\documentclass[a4paper,11pt]{article}
\pdfoutput=1
\usepackage{jheppub}
\usepackage[dvipsnames]{xcolor}
\usepackage[utf8]{inputenc}
\usepackage[T5]{fontenc}
\usepackage[english]{babel}
\usepackage{mathtools}
\usepackage{amsfonts}
\usepackage{amsthm}
\usepackage{physics}
\usepackage{comment}
\usepackage{float}
\usepackage{subcaption}
\usepackage{tikz}
\usepackage{tkz-euclide}
\usetikzlibrary{decorations.markings}
\usetikzlibrary{patterns}
\usepackage{tikz-feynman}
\usepackage{simpler-wick}
\usepackage{booktabs} 
\usepackage{hyperref}
\hypersetup{urlcolor=magenta, citecolor=Maroon, linkcolor=NavyBlue}
\usepackage{url}

\allowdisplaybreaks
\renewcommand{\a}{\alpha}

\newcommand{\g}{\gamma}

\newcommand{\s}{\sigma}
\renewcommand{\t}{\tau}
\newcommand{\D}{\Delta}

\newcommand{\G}{\Gamma}
\renewcommand{\mathbf}[1]{\textbf{#1}} 
\newcommand{\DeclareAutoPairedDelimiter}[3]{%
  \expandafter\DeclarePairedDelimiter\csname Auto\string#1\endcsname{#2}{#3}%
  \begingroup\edef\x{\endgroup
    \noexpand\DeclareRobustCommand{\noexpand#1}{%
      \expandafter\noexpand\csname Auto\string#1\endcsname*}}%
  \x}
\DeclareAutoPairedDelimiter{\p}{(}{)}

\usepackage{bbm}
\usepackage{bm} 
\renewcommand{\it}[1]{\textit{#1}} 

\usepackage{cancel} 

\renewcommand{\tilde}[1]{\widetilde{#1}}
\DeclarePairedDelimiterX{\avg}[1]{\langle}{\rangle}{#1} 

\renewcommand{\L}{\mathcal{L}}

\theoremstyle{plain}

\author{Ziyi Li}
\affiliation{Department of Physics, University of California at Davis, CA 95616, U.S.A.}
\emailAdd{zzyli@ucdavis.edu}

\title{Spinning States and Unitarity in 3D Gravity}

\abstract{We revisit the proposal to cure the negative density of states in the three-dimensional gravitational path integral by adding spinning states whose spin scales with the central charge. We show that sub-extremal and extremal spinning states below the black hole threshold can cancel the known negativities, and interpret these states as bulk spinning defects. Additionally, certain overspinning states above the black hole threshold can cure these negativities while preserving the spectral gap. The overspinning states were previously interpreted as classical spinning strings; we instead identify them as overspinning BTZ geometries, which are smooth pure gravity quotients of AdS$_3$ with no fixed points. All of these spinning geometries exhibit causal pathologies in their Lorentzian continuations. Moreover, the overspinning geometries arise from mixed elliptic-hyperbolic identifications and contain a formally real chiral temperature. We also generalize the computation of scalar correlators to the extremal background and propose a quotient-motivated image expression for the overspinning geometries.}
\newpage

\begin{document} 
\maketitle
\flushbottom

\section{Introduction}
Gravity in (2+1) dimensions with a negative cosmological constant has proven to be a useful arena for exploring toy models of quantum gravity. Although the theory has no local propagating degrees of freedom, it does contain black hole solutions \cite{Banados:1992wn,Banados:1992gq}, whose existence is certainly not a coincidence. As a precursor of the AdS/CFT duality \cite{Maldacena:1997re,Witten:1998qj,Gubser:1998bc}, Brown and Henneaux demonstrated that the asymptotic symmetry of the theory with suitable boundary conditions is the infinite dimensional Virasoro symmetry, with central charge \cite{Brown:1986nw}: 
\begin{align}
\label{bhcentralcharge}
    c=\frac{3l}{2G}
\end{align}
where $l$ is the AdS$_3$ radius. This highly constrains the theory, and the asymptotic density of states can be computed using Cardy's formula \cite{Cardy:1986ie,Bloete:1986qm}, which precisely matches the BTZ black hole entropy \cite{Strominger:1997eq,Birmingham:1998jt}. Although this can be viewed as a microscopic counting of the BTZ black hole microstates, the argument that leads to Cardy's formula relies on modular invariance, and the details of these microstates are still lacking. Nonetheless, this strongly suggests that we should take the existence of the black hole in (2+1) dimensions seriously. From a modern viewpoint, solving three-dimensional quantum gravity with a negative cosmological constant amounts to finding a two dimensional conformal field theory with the central charge (\ref{bhcentralcharge}). However, no such theory has been definitively found \cite{Witten:2007kt,Yin:2007gv,Giombi:2008vd}.

Despite not knowing the exact dual conformal field theory, one can still compute the partition function of pure three-dimensional gravity by summing over all classified Euclidean saddles \cite{Dijkgraaf:2000fq} in the gravitational path integral \cite{Maloney:2007ud,Keller:2014xba}, which we review in appendix \ref{appenA}. The resulting sum is given in terms of Poincar\'e series, and the density of states is known to contain several unphysical features. First, the spectrum is continuous rather than discrete, which challenges the conventional interpretation of pure three-dimensional gravity being dual to a single CFT$_2$ \cite{Chandra:2022bqq}. More severely, the spectrum exhibits negative density of states in two regimes: at the BTZ black hole threshold \cite{Maloney:2007ud} and at large-spin $\abs{j}\to \infty$ near-extremality \cite{Benjamin:2019stq}. The negative density of states renders the spectrum non-unitary, which might suggest that pure three-dimensional quantum gravity does not exist.

Nonetheless, many attempts have been made to render the spectrum positive definite. The negativity at the scalar black hole threshold can be rendered positive by adding six compact boson partition functions \cite{Keller:2014xba}, and the negativity at large-spin can be canceled by adding scalar states that scale as the central charge below the black hole threshold \cite{Benjamin:2019stq}. In particular, \cite{Benjamin:2020mfz} proposed adding a finite number of defect states that lie below the threshold. These conical defects have a natural geometric interpretation as heavy point particles from the early days of three-dimensional gravity \cite{Deser:1983nh,Deser:1983tn}. These geometries have vanishing stress-energy tensor everywhere except at the location of the singularity, which deviates from purely metric degrees of freedom. Another proposal that preserves the pure-gravity spectrum while curing the large-spin negativity is to include off-shell geometries known as Seifert manifolds \cite{Maxfield:2020ale}. By dimensionally reducing to JT gravity, the addition of conical defects (Seifert manifolds in 3D) non-perturbatively shifts the BTZ extremality bound, rendering the spectrum positive, while still leaving the threshold negativity intact. One way to cure both the threshold negativity and the large-spin negativity was proposed in \cite{DiUbaldo:2023hkc}, where the authors added two heavy states that sit exactly at the BTZ threshold, with spin scaling linearly with the central charge, canceling both the large-spin and threshold negativities, while requiring $\xi=\frac{c-1}{24}\in 2\mathbb{Z}_+$. In fact, the observation that adding states with spin can cancel both regimes of negativity was already anticipated in \cite{Benjamin:2020mfz}, where the authors proposed adding heavy states below the BTZ threshold with $\mathcal{O}(1)$ spin. Geometrically, the spinning states in \cite{DiUbaldo:2023hkc} have zero BTZ black hole mass and non-zero angular momentum, rendering them overspinning. The authors interpreted these states as the classical spinning strings \cite{Maxfield:2022rry}, whose metric outside of the string is given by the overspinning BTZ geometries \cite{Briceno:2021dpi}. 

In this paper, we instead study the realization of these overspinning states as overspinning BTZ geometries, whose quotient spacetime is smooth and free of singularities. The Lorentzian quotient is smooth and free of fixed points, which provides a realization of the same asymptotic charges without an explicit matter source. However, it does contain closed timelike curves and a second asymptotic end. A further distinctive feature is that overspinning geometries arise from identifications of mixed elliptic and hyperbolic generators, unlike their sub-extremal counterparts, which are purely elliptic. As a result, the hyperbolic sector carries formally a real temperature, and we refer to these backgrounds as \textit{partially thermal}. Due to the elliptic-hyperbolic identifications, adopting the standard method of computing the correlator we use for (sub)-extremal defects will require an analytic continuation of the metric into the complex plane, and imposing boundary conditions at complex radial values. We will instead propose a quotient motivated image expression for the correlator, whose leading contribution matches the analytically continued CFT$_2$ vacuum block. In addition, we also show that certain overspinning states above the BTZ threshold can similarly cure the negativities discussed above, preserving the spectral gap while requiring $\frac{2}{3}\xi \in \mathbb{Z}_+$. 

In addition to the overspinning states, we show that certain sub-extremal and extremal spinning states can cure the same negativities in the regimes discussed above. Adding these geometries will impose a stricter constraint on the central charge, requiring $\xi\in5 \mathbb{Z}_+$ and $\xi \in 4 \mathbb{Z}_+$ for spinning and extremal states respectively. We interpret these spinning states as spinning defects with a localized source and a surrounding region of CTCs. Their scalar correlators can be studied with the standard quantization method, although the full-range normalization integral still crosses the chronology-violating region and should be understood as an analytic prescription. 

These spectral constructions and Lorentzian geometries do not by themselves establish additional Euclidean saddles of the gravitational path integral. Although Lorentzian causal pathologies
alone need not exclude a Euclidean saddle, the quasi-Euclidean continuation of the overspinning metric fails the Kontsevich–Segal–Witten criterion \cite{Kontsevich:2021dmb,Witten:2021nzp,Basile:2023ycy}, as well as the sub-extremal and extremal defects retaining the CTC region. Due to the strict nature of the KSW criterion, we still would like to regard our candidate bulk interpretations as reasonable saddles whose contributions to the Euclidean path integral remains to be understood.

\paragraph{Outline.}
The outline of the paper is as follows. In section~\ref{section2}, we study the addition of sub-extremal, extremal, and overspinning states to the MWK partition function. We prove the scalar sector positivity for the sub-extremal and extremal families and analyze the nonzero-spin sectors at leading semiclassical order. In section \ref{section3}, we discuss the proposed bulk interpretations for these states. We adopt the spinning defect completion for the sub-extremal and extremal states, and propose the maximal-quotient interpretation for the overspinning states. We discuss their causal properties and scalar correlators, as well as the KSW criterion for the quasi-Euclidean continuation of these geometries and its possible implications. We end with discussions in section \ref{section4}.

\section{Spinning states and MWK unitarity}
\label{section2}
The MWK partition function \cite{Maloney:2007ud,Keller:2014xba} is constructed by summing over the classified Euclidean saddles \cite{Dijkgraaf:2000fq} in the gravitational path integral, which are thermal AdS$_3$ and $SL(2,\mathbb{Z})$ black holes. This can be viewed as summing over $PSL(2,\mathbb{Z})$ images of the Virasoro vacuum character in the dual CFT$_2$: 
\begin{align}
\label{mkw}
     Z_{\text{MWK}}(\tau,\bar{\tau})=
    \sum _{\g \in \Gamma_\infty\backslash PSL(2,\mathbb{Z})}
    \abs{\chi_{\text{vac}}\p{\g \tau}}^2.
\end{align}
Despite being such a simple prescription, there are several problems with this sum. First, the spectrum is continuous rather than discrete, which challenges the conventional interpretation of pure three-dimensional gravity being dual to a single CFT$_2$ \cite{Chandra:2022bqq}. More severely, the regularized spectrum exhibits negative density of states in two regimes: at large-spin $\abs{j}\to \infty$ near extremality, and at the BTZ black hole threshold. At the scalar black hole threshold, the regularized density of states contains a $-6$ degeneracy plus a continuous function: 
\begin{align}
    \tilde{\rho}^{\text{MWK}}_0(t)=-6\delta(t)+\frac{2}{t}\sum_{s=1}^{\infty} \biggl\{\frac{\phi(s)}{s}\left[\sinh^2\p{\frac{4\pi}{s}\sqrt{\xi t}}+\sinh^2\p{\frac{4\pi}{s}\sqrt{(\xi-1)t}}\right] \notag \\
    -2\frac{\mu(s)}{s}\left[\cosh\p{\frac{4\pi}{s}\sqrt{\xi t}}\cosh\p{\frac{4\pi}{s}\sqrt{(\xi-1)t}}-1\right]\biggl\}
\end{align}
where $t=\text{min}(h,\bar h)-\xi$ is the reduced twist, and $\xi=\frac{c-1}{24}$. We use $j=h-\bar h$ and $\bar t=t+\abs{j}$ for the continuum, and $J=H-\bar H$, $T=\min(H,\bar H)-\xi$, $\bar T=T+\abs{J}$ for a seed. Here $\phi(s)$ and $\mu(s)$ are Euler's totient and M\"obius functions. It was first proposed in \cite{Keller:2014xba} that the leading negative contribution can be canceled by adding six compact bosons. Later, \cite{Benjamin:2020mfz} argued that the same term can instead be removed by adding states with spin of order $\mathcal{O}(1)$. More recently, \cite{DiUbaldo:2023hkc} showed explicitly that the negativity can be cured by including states whose spin scales with the central charge, $J\sim \mathcal{O}(c)$. The proposal of adding states with spin to cancel the above negativity stems from the observation that the regularized $j=0$ sum for a seed state $T<0,J>0$ is given by \cite{Maloney:2007ud,Benjamin:2020mfz}: 
\begin{align}
\label{seedwithzerospin}
    \tilde{\rho}_{j=0}^{(T,J)}(t)=2\sigma_0(J)\delta(t)+\sum_{s=1}^{\infty}\frac{2S(0,J;s)}{st}\left[\cosh\p{\frac{4\pi}{s}\sqrt{-Tt}}\cosh\p{\frac{4\pi}{s}\sqrt{-\bar{T}t}}-1\right]
\end{align}
where $\sigma_0(\abs{J})$ is the divisor-counting function. Note that $\sigma_0(\abs{J})\geq1$ for every nonzero integer $J$. This imposes a condition on $\xi$ if $J \propto \xi$. Thus, by adding certain numbers of these spinning states, the negativity at the threshold can be canceled. For the overspinning states, the continuous piece in (\ref{seedwithzerospin}) can in fact introduce new negativity in the regime $\xi t\lesssim \mathcal{O}(1)$, and ensuring this positivity will place a bound on the degeneracy of spinning states one can add to the partition function \cite{DiUbaldo:2023hkc}. This restriction is no longer present for sub-extremal and extremal spinning states below the threshold.

Another regime of negativity in the MWK partition function appears in the large-spin and near-extremal limit ($j \to \infty$, $t\to 0$), where the density of states is given by \cite{Benjamin:2019stq}: 
\begin{align}
    \rho^{\text{MWK}}_j(t)\sim \frac{8\pi^2}{\sqrt{j}}\sqrt{t} \hspace{1mm} e^{4\pi \sqrt{\xi j}}+\sum_{s=2}^\infty &\left[\frac{S(j,0;s)-S(j,-1;s)}{s\sqrt{tj}}e^{\frac{4\pi}{s}\sqrt{\xi j}} \right.  \notag \\
   &+ \left. \frac{S(j,0;s)-S(j,1;s)}{s\sqrt{tj}}e^{\frac{4\pi}{s}\sqrt{\p{\xi-1} j}} \right]
\end{align}
In the near-extremal limit, the second and third terms dominate, and the density of states can become negative, originating from the difference of Kloosterman sums. Since higher $s$ terms will be exponentially suppressed, we can focus on the $s=2$ term which causes the leading negativity. Using $S(j,0;2)=(-1)^j=-S(j,-1;2)=-S(j,1;2)$, the $s=2$ term yields in the limit $j\to \infty$: 
\begin{align}
\label{MWK Large Spin}
    \rho^{\text{MWK}}_j(t)\sim \frac{(-1)^j}{\sqrt{tj}}\p{e^{2\pi\sqrt{\xi j}}+e^{2\pi\sqrt{\p{\xi-1} j}}}
\end{align}
Indeed, for odd spins $j$, the density of states becomes negative with exponential behavior $e^{2\pi\sqrt{\xi j}}$. To cancel this negativity, it was observed that a seed state with reduced twist
\begin{align}
    T=\text{min}\p{H,\bar H}-\xi=-\frac{1}{4}\xi
\end{align}
has the same exponential behavior in this large-spin limit but with a positive prefactor, and adding certain numbers of such states can indeed cure the large-spin negativity. This was the route taken by \cite{Benjamin:2019stq,Benjamin:2020mfz,DiUbaldo:2023hkc}. In particular, \cite{Benjamin:2019stq} considered adding a finite number of scalar states with weight $H=\bar H=\frac{c-1}{32}$, while \cite{Benjamin:2020mfz} considered the minimum particle content required to cancel the above negativity as well as the negativity from subleading $s$ terms, and interpreted these states as conical defects in the bulk. These proposals of adding scalar states only address the large-spin negativity, and the negativity at the threshold can only be canceled by adding boson partition functions. To cancel both regimes of negativity at once, we need to add states with spin. This is indeed the proposal of \cite{DiUbaldo:2023hkc}, where, by adding two spin-parity seed states each with a degeneracy of two, they were able to cancel the negativity in both regimes. In particular, the seed states are given by: 
\begin{align}
    (H,\bar{H})=(\frac{5}{4}\xi,\frac{3}{4}\xi)
\end{align}
designed to cancel the $j\to -\infty$ negativity together with the spin-parity state $H \leftrightarrow\bar H$, which cures the $j \to \infty$ negativity. In terms of the bulk mass $\mathbf{M}$ and spin $\mathbf{J}$, these states have, at leading semiclassical order,\footnote{The parameters $\mathbf{M},\mathbf{J}$ appear in the BTZ metric, with AdS radius set to one. With $c=3/(2G)$, the charge relations are $H+\bar H=(\mathbf{M}+1)/(8G)=\mathbf{M}/(8G)+2\xi+1/12$ and $H-\bar H=\mathbf{J}/(8G)$. Thus, the mass-zero threshold of the classical metric is at $H+\bar H=c/12$, whereas the shifted primary threshold is at $2\xi=(c-1)/12$. The displayed bulk charges omit $O(1/c)$ corrections.}
\begin{align}
    \mathbf{M}\simeq0, \quad \mathbf{J}\simeq\pm \frac{1}{4}.
\end{align}
These seeds lie exactly at the shifted primary-dimension threshold, preserving the dimension gap while carrying nonzero spin. In this semi-classical sense, they are overspinning, satisfying $\abs{\mathbf{M}}<\abs{\mathbf{J}}$. In \cite{DiUbaldo:2023hkc}, the authors interpreted these states as classical spinning strings \cite{Maxfield:2022rry}, introducing matter into the pure gravity spectrum. In section \ref{section3}, we instead study the maximal overspinning BTZ quotient as a source-free Lorentzian completion with the same asymptotic charges. The Lorentzian geometry is smooth but contains CTCs and a second asymptotic region. However, its inclusion in a one-boundary Euclidean path integral is not established by these Lorentzian properties.

Another possibility that we will explore in this paper is to add sub-extremal spinning and extremal states to the path integral, since these geometries are less ambiguous in terms of their bulk interpretation but nonetheless contain causal pathologies in their Lorentzian geometries. Before turning to the details, we summarize the main results in Table \ref{summarizedresults}.

\begin{table}[ht]
\centering
\renewcommand{\arraystretch}{1.2}
\small
\setlength{\tabcolsep}{7pt}
\begin{tabular}{@{} l c c c c @{}}
\toprule
Type & $H$ & $\bar H$ & $\xi$ & Scalar-sector result \\
\midrule
Sub-extremal &   $\frac{19}{20}\xi$   &  $\frac{3}{4}\xi$  & $\xi \in 5\mathbb{Z}_+$ & No upper bound on $d$\\
Extremal & $\xi$ & $\frac{3}{4}\xi$ &  $\xi \in 4\mathbb{Z}_+$    & No upper bound on $d$ \\
Overspinning & $\frac{17}{12}\xi$  & $\frac{3}{4}\xi$ & $\xi \in \frac{3}{2}\mathbb{Z}_+$ & $d=2$ sufficient \\
\bottomrule
\end{tabular}
\caption{This table summarizes our results: Adding any of the above seed states together with its spin-parity state $J<0$ ($H \leftrightarrow \bar H$), each with a degeneracy $d=2$, is sufficient to cancel the known MWK negativities without introducing new ones. In particular, any $H \in \left(\frac{3}{4}\xi,\, \xi\right)$ with $\bar{H}=\frac{3}{4}\xi$ can achieve the same cancellation; the spin quantization $J = H - \bar{H} \in \mathbb{Z}$ then constrains $\xi$. We choose $(H,\bar{H}) = \left(\frac{19}{20}, \frac{3}{4}\right)\xi$, which yields $\xi \in 5\mathbb{Z}_+$, the least restrictive quantization condition. The overspinning states preserve the spectral gap of $\frac{c-1}{12}$, while (sub)-extremal spinning states are below the black hole threshold. }
\label{summarizedresults}
\end{table}

\subsection{Sub-extremal spinning states}
The sub-extremal spinning states lie below the black hole threshold, and they can be viewed as created by a heavy non-scalar primary operator in the dual CFT$_2$. By sub-extremal, we simply mean that the state has $\abs{\mathbf{M}}>\abs{\mathbf{J}}$, and in terms of the conformal dimensions this can be expressed as: 
\begin{align}
\label{underspinning}
 \bar T=  \text{max}\p{H,\bar H}-\xi < 0
\end{align}
Following the construction in \cite{DiUbaldo:2023hkc}, we add two seed spin-parity states designed to cancel the $j\to \pm \infty$ negativity. Moreover, we choose each state to have a degeneracy $d=2$ so that the first term in (\ref{seedwithzerospin}) renders the threshold $-6\delta(t)$ term nonnegative. Using (\ref{seedwithzerospin}) we have for the delta function piece: 
\begin{align}
    2d \cdot \tilde{\rho}_{j=0}^{(T,J)}(t)\supset 4d\sigma_0(J)\delta(t)
\end{align}
where the factor of two comes from the seed state and its parity image. Including the MWK term, the coefficient is thus $4d\sigma_0(\abs{J})-6>0$ for $d=2$.

To cancel the large-spin negativities, we need to reproduce the same leading behavior in the same regime as in equation (\ref{MWK Large Spin}), which requires our seed state to have: 
\begin{align}
    \text{min}\p{H,\bar H}=\frac{3}{4} \xi
\end{align}
This follows from the density of states obtained by the $PSL(2,\mathbb{Z})$ sum over a seed state of reduced twist $T<0$ and spin $J$, which for $jJ\geq 0$ takes the form \cite{Benjamin:2020mfz}: 
\begin{align}
\label{seed with spin}
    \rho_j^{\p{T,J}}(t)=\frac{2}{\sqrt{t \bar{t}}}\sum_{s=1}^\infty\frac{1}{s}S(j,J,s)\cosh\p{\frac{4\pi}{s}\sqrt{-\bar{T}\bar{t}}}\cosh\p{\frac{4\pi}{s}\sqrt{-Tt}}
\end{align}
while for $jJ\leq 0$, we have: 
\begin{align}
\label{seed with spin2}
    \rho_j^{\p{T,J}}(t)=\frac{2}{\sqrt{t \bar{t}}}\sum_{s=1}^\infty\frac{1}{s}S(j,J,s)\cosh\p{\frac{4\pi}{s}\sqrt{-\bar{T}t}}\cosh\p{\frac{4\pi}{s}\sqrt{-T\bar{t}}}
\end{align}
In the large-spin and near-extremal limit of (\ref{seed with spin2}), where we assume that $j\to -\infty$, $t\to0$ with $J>0$, the leading $s=1$ term yields: 
\begin{align}
    \rho_j^{(T,J)} \sim \frac{d}{\sqrt{-tj}}e^{4\pi \sqrt{\p{-\bar{t}\hspace{0.5mm}T}}}
\end{align}
where $T=\bar{H}-\xi$ ($J>0$) and $\bar{t}=-j>0$. Requiring the above density of states to scale like (\ref{MWK Large Spin}) yields $T=-\frac{1}{4}\xi$, which translates to:
\begin{align}
   -( \bar{H}-\xi )=\frac{1}{4}\xi \implies \bar{H}=\frac{3}{4}\xi
\end{align}
which cancels the $j\to -\infty$ negativity. The $j\to \infty$ negativity can then be canceled by the spin-parity state with $H=\frac{3}{4}\xi$ ($J<0$). The leading scaling only fixes the reduced twist $T$, and it appears that we have complete freedom in choosing $\bar T$ within the range:
\begin{align}
\label{boundsubex}
    \frac{3}{4}\xi<\text{max}\p{H,\bar H}<\xi
\end{align}
where the lower bound comes from the condition we impose on $T$, and the upper bound is the requirement for the state to be sub-extremal. However, the quantization condition for $J$ will be different, and this imposes different conditions on the central charge $\xi$. A convenient rational subfamily has $J=\xi/n$ with $n\in\mathbb Z_{\geq5}$ and the condition on the central charge is $\xi\in n\mathbb Z_+$. Since we do not want to impose too strict a condition on $\xi$, we choose smallest allowed denominator: 
\begin{align}
    (H,\bar{H})=\left(\frac{19}{20},\frac{3}{4}\right)\xi
\end{align}
This gives $J=\frac{1}{5}\xi$, and $\xi \in 5\mathbb{Z}_+$, which is a sparser arithmetic condition than the $\xi \in 2\mathbb{Z}_+$ of \cite{DiUbaldo:2023hkc}. The total density of states for the sub-extremal states is then given by: 
\begin{align}
\label{subextotal}
    \rho^{\text{sub-ext}}_j=&\frac{2d}{\sqrt{t\bar{t}}}\sum_{s=1}^\infty \frac{1}{s}S(j,J;s) \cosh\p{\frac{2\pi}{s}\frac{1}{\sqrt{5}}\sqrt{\xi \bar t}}\cosh \p{\frac{2\pi}{s}\sqrt{\xi t}}\notag \\+&\frac{2d}{\sqrt{t\bar{t}}}\sum_{s=1}^\infty \frac{1}{s}S(j,-J;s) \cosh\p{\frac{2\pi}{s}\frac{1}{\sqrt{5}}\sqrt{\xi  t}}\cosh \p{\frac{2\pi}{s}\sqrt{\xi \bar t}}
\end{align}
where we choose $d=2$. In addition to curing the existing negativities, we also need to ensure that the above density of states does not introduce new negativities, which we turn to next. 

\subsubsection{Positivity}
We need to demonstrate positivity for the new total density of states given by: 
\begin{align}
    \rho_j^{\text{total}}= \rho_j^{\text{MWK}}+\rho_j^{\text{sub-ext}}
\end{align}
in various regimes \cite{DiUbaldo:2023hkc}. The parity renders the density even in $j$, so we focus on establishing positivity for $j\geq1$, both at finite $t$ and in the limit $t\to 0$, assuming $\xi\gg1$. In the extremal ($t \to 0$) regime, we are still assuming $\xi \gg 1$ even for finite $j$. Thus, the argument for large $j$ limit carries through, and positivity is established. At finite $t$, we still have $\xi j\gg1$, and the MWK density of states (\ref{MWK partition function}) is dominated by the positive leading $s=1$ term, which scales like $e^{4\pi \sqrt{\xi}\p{\sqrt{t}+\sqrt{t+j}}}$. Similarly, the total sub-extremal density of states is dominated by the $s=1$ term, which scales like $e^{2\pi \sqrt{\xi \bar t}}$ with a subleading piece $e^{\frac{2\pi}{\sqrt{5}}\sqrt{\xi t}}$. Thus, the positivity of the total density of states is established for $j\geq1$.

\begin{figure}[h]
    \centering
    \begin{minipage}{0.49\textwidth}
        \centering
        \includegraphics[width=\linewidth]{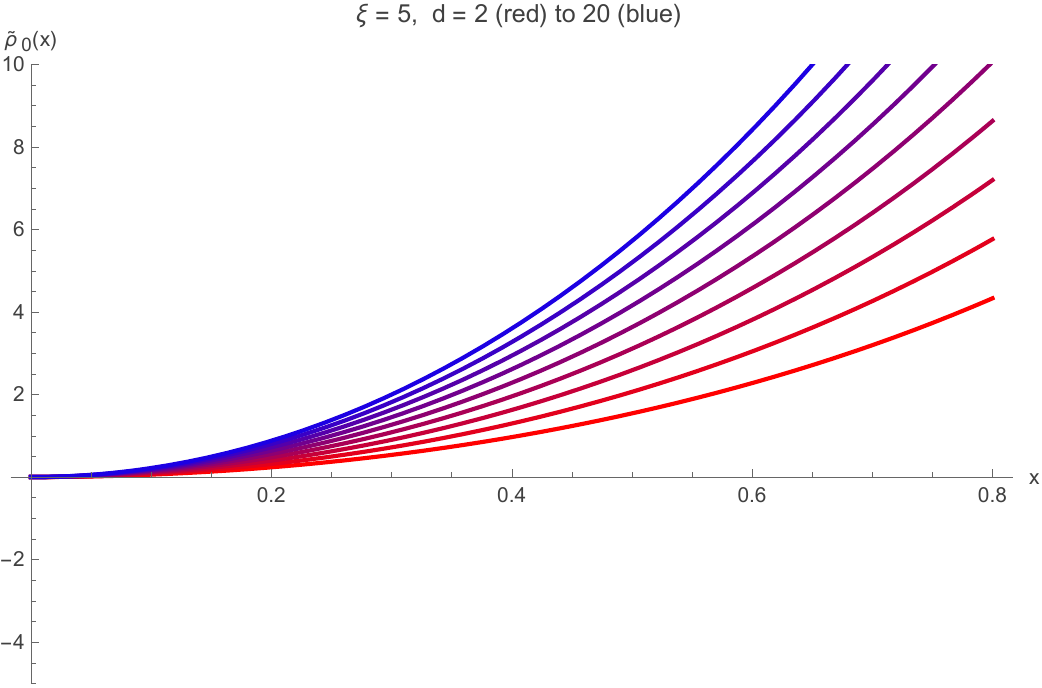}
    \end{minipage}
    \hfill
    \begin{minipage}{0.48\textwidth}
        \centering
        \includegraphics[width=\linewidth]{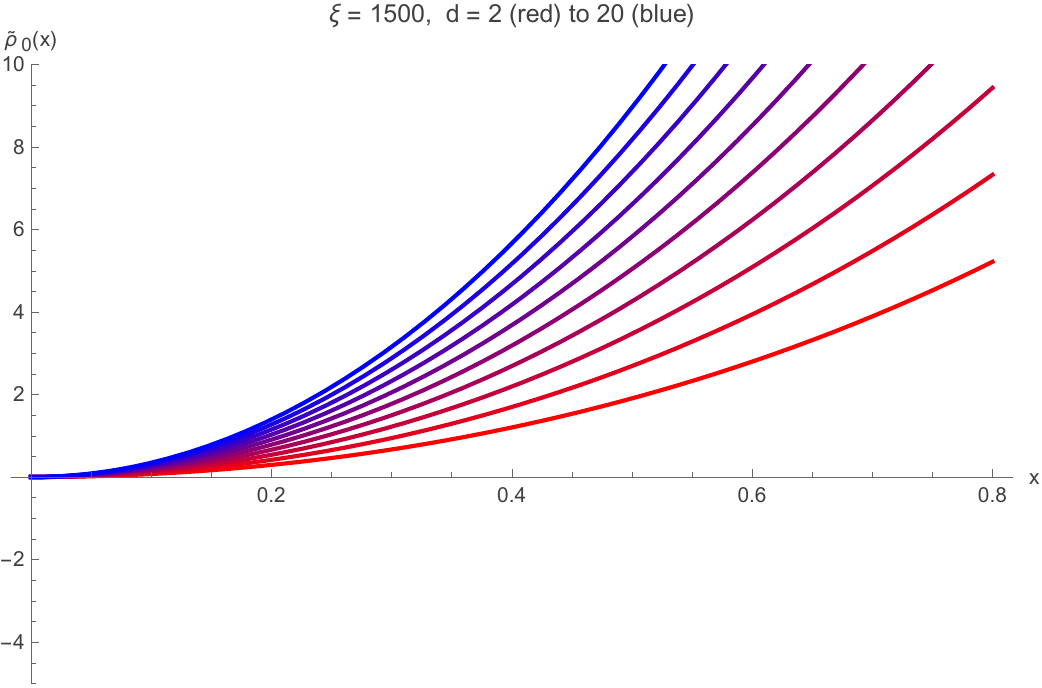}
    \end{minipage}
    \caption{The total scalar density ($\frac{t}{2}\tilde{\rho}_0^{\hspace{1mm}\text{total}}$) obtained after adding sub-extremal spinning states. The left graph corresponds to $\xi=5$, the smallest allowed value, with $d$ ranging from $2$ (red) to $20$ (blue) in increments of $2$, while the right graph shows the analogous result for $\xi=1500$. No negative region develops in either case. The sum becomes more positive as $d$ increases. Both plots are obtained by summing over $s\leq 300$.}
    \label{fig:subext}
\end{figure}

Now we turn to the positivity of the continuous part of the scalar density at $j=0$. Since the functions involved are all hyperbolic, the positivity is more evident than in the overspinning case. The continuous part of the MWK scalar density is given by: 
\begin{align}
     \frac{t}{2}\tilde\rho^{\mathrm{MWK}}_{0,\mathrm{cont}}(t)
 =\sum_{s=1}^{\infty}\frac{1}{s}  \left\{\phi(s)[\sinh^2 A_s+\sinh^2 B_s]-2\mu(s)[\cosh A_s\cosh B_s-1]\right\}
\end{align}
where we have defined $A_s=4\pi\sqrt{\xi t}/s$ and $B_s=4\pi\sqrt{(\xi-1)t}/s$. For $\xi\geq 1$, every term in the summation is nonnegative. This is obvious for $\mu(s)\leq0$. For $\mu(s)=1$, we can use $\phi(s)\geq1$ to obtain the following positivity bound: 
\begin{align}
 \sinh^2 A_s+\sinh^2 B_s-2[\cosh A_s\cosh B_s-1] =(\cosh A_s-\cosh B_s)^2\geq0.
\end{align}
Next, the scalar density of states of the sub-extremal spinning state is given by: 
\begin{align}
    \tilde{\rho}_0^{\text{sub-ext}}(t)=\frac{4d}{t}\sum_{s=1}^{\infty} \frac{c_s(J)}{s}\left[\cosh\p{\frac{2\pi}{s}\frac{1}{\sqrt{5}}\sqrt{\xi t}}\cosh\p{\frac{2\pi}{s}\sqrt{\xi t}}-1\right]
\end{align}
The hyperbolic functions above admit the following Taylor series expansion, with positive coefficients:
\begin{align}
\cosh&\left(\frac{x}{\sqrt5\,s}\right)\cosh\left(\frac{x}{s}\right)-1
= \sum_{n=1}^{\infty} a_n \frac{x^{2n}}{s^{2n}},  \hspace{0.8cm}a_n = \sum_{m=0}^{n} \frac{1}{(2m)!\hspace{0.6mm}(2(n-m))!\hspace{0.4mm} 5^m}>0, 
\end{align}
where we have defined $x=2\pi \sqrt{\xi t}$ and $c_s(J)=S(0,J;s)$ is a Ramanujan sum. Plugging this back into the $s$ sum yields: 
\begin{align}
 \tilde{\rho}_0^{\text{sub-ext}}(t)
= \frac{4d}{t} \sum_{n=1}^{\infty} a_n x^{2n}
\sum_{s=1}^{\infty}\frac{c_s(J)}{s^{2n+1}}.
\end{align}
Since the original sum is absolutely convergent, we have freely rearranged terms and interchanged the $s$ sum with the power series in $x$. Performing the sum over $s$ using the standard Dirichlet series identity for Ramanujan sums gives: 
\begin{align}
\sum_{s=1}^{\infty}\frac{c_s(J)}{s^{2n+1}}
=
\frac{\sigma_{-2n}(J)}{\zeta(2n+1)}.
\end{align}
Since $\sigma_{-2n}(J)>0$ and $\zeta(2n+1)>0$, every coefficient in the expansion of $\tilde{\rho}_0^{\hspace{1mm}\text{sub-ext}}(t)$ is strictly positive for all $x$. Thus, adding these sub-extremal spinning states will in fact introduce no new negativities, and there is no upper bound on the degeneracy of the state $d$. This is also confirmed numerically: as shown in Fig. \ref{fig:subext}, the total scalar density becomes increasingly positive with larger $d$.

\subsection{Extremal spinning state }
\begin{figure}[h]
    \centering
    \begin{minipage}{0.49\textwidth}
        \centering
        \includegraphics[width=\linewidth]{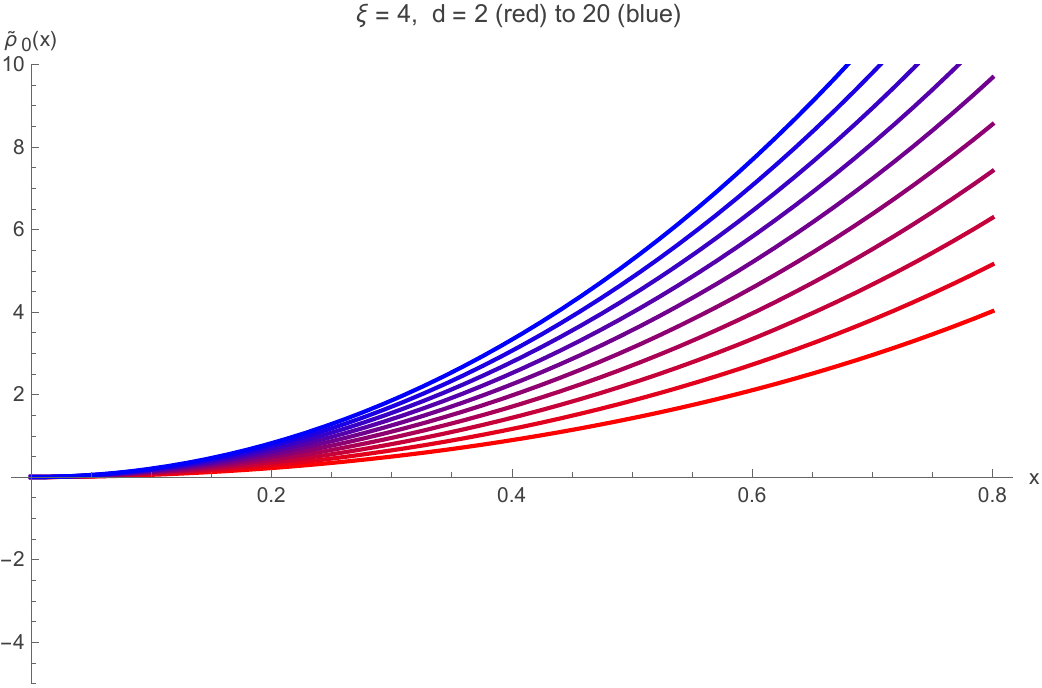}
    \end{minipage}
    \hfill
    \begin{minipage}{0.48\textwidth}
        \centering
        \includegraphics[width=\linewidth]{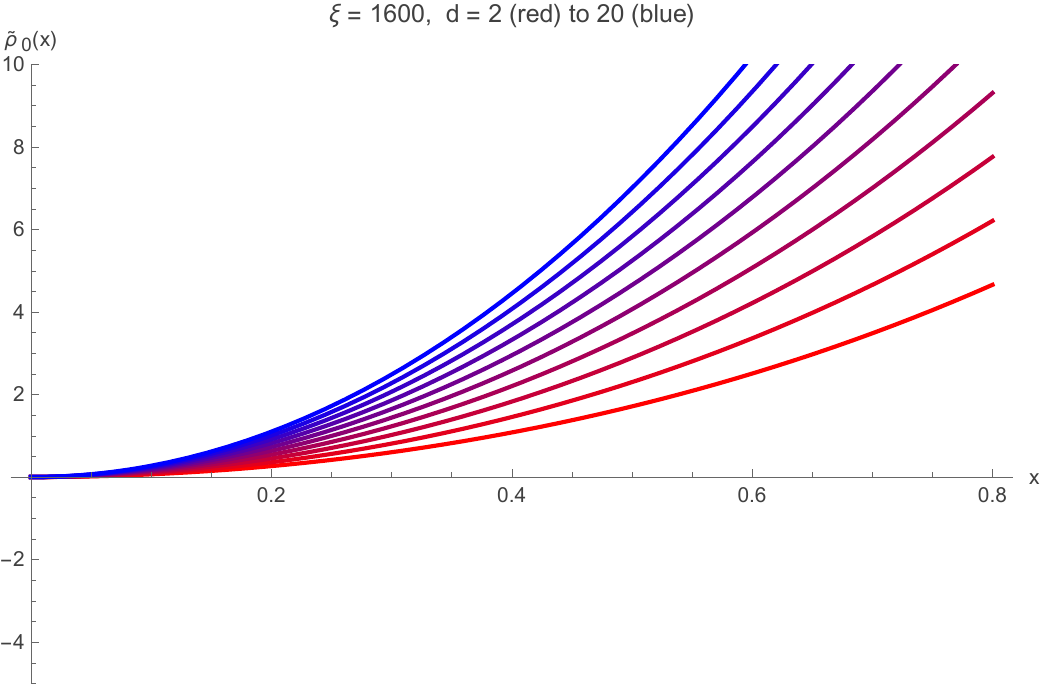}
    \end{minipage}
    \caption{The total scalar density obtained after adding extremal spinning states. The left graph corresponds to $\xi=4$, the smallest allowed value, with $d$ ranging from $2$ (red) to $20$ (blue) in increments of $2$, while the right graph shows the analogous result for $\xi=1600$.}
    \label{fig:eext}
\end{figure}

The extremal state is given by: 
\begin{align}
    \text{max}\p{H,\bar H}=\xi \implies \bar T=0
\end{align}
By extremal, we simply mean that the bulk parameters satisfy $\abs{\mathbf{M}}=\abs{\mathbf{J}}$ in the semiclassical limit. Again we need to fix $T=-\frac{1}{4}\xi$ to cancel the large-spin negativity. Thus, the proposal here is to add two spin-parity states given by: 
\begin{align}
     (H,\bar{H})=\left(1,\frac{3}{4}\right)\xi
\end{align}
each with degeneracy $d=2$, together with the condition $\xi \in 4\mathbb{Z}_+$. The density of states of the extremal spinning states is given by: 
\begin{align}
\label{extotal}
    \rho^{\text{ext}}_j=&\frac{2d}{\sqrt{t\bar{t}}}\sum_{s=1}^\infty \frac{1}{s}S(j,J;s) \cosh \p{\frac{2\pi}{s}\sqrt{\xi t}} +\frac{2d}{\sqrt{t\bar{t}}}\sum_{s=1}^\infty \frac{1}{s}S(j,-J;s) \cosh \p{\frac{2\pi}{s}\sqrt{\xi \bar t}}
\end{align}
The positivity of the total density of states for $j\geq 1$ follows in the same way as in the sub-extremal case, both for finite $t$ and in the limit $t\to 0$, with $\xi\gg 1$. The proof for the scalar density also proceeds analogously. With $x=2\pi\sqrt{\xi t}$, we have
\begin{align}
 \tilde\rho_{0,\mathrm{cont}}^{\mathrm{ext}}(t)
 &=\frac{4d}{t}\sum_{s=1}^{\infty}\frac{c_s(J)}{s}
 [\cosh(x/s)-1]\notag\\
 &=\frac{4d}{t}\sum_{n=1}^{\infty}\frac{x^{2n}}{(2n)!}
 \frac{\sigma_{-2n}(J)}{\zeta(2n+1)}>0.
\end{align}
Thus, the extremal pair also places no scalar-sector upper bound on $d$. We can also plot the scalar density for the extremal spinning state, and observe the same behavior, as shown in Fig. \ref{fig:eext}.

\subsection{Overspinning states above BTZ threshold}
\begin{figure}[h]
    \centering
    \begin{minipage}{0.49\textwidth}
        \centering
        \includegraphics[width=\linewidth]{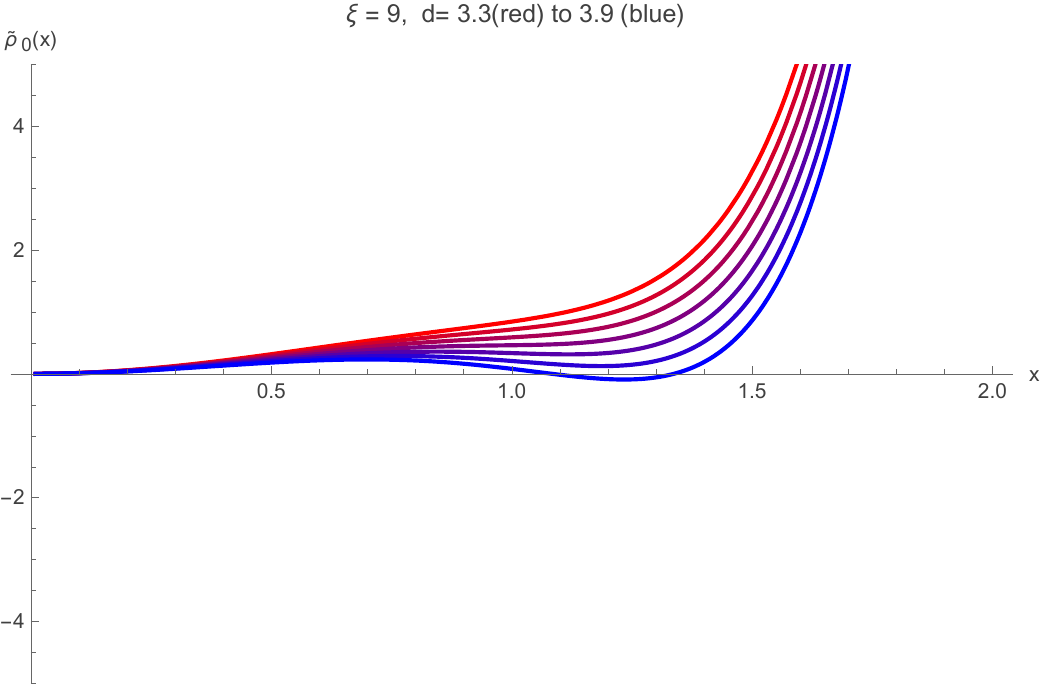}
    \end{minipage}
    \hfill
    \begin{minipage}{0.48\textwidth}
        \centering
        \includegraphics[width=\linewidth]{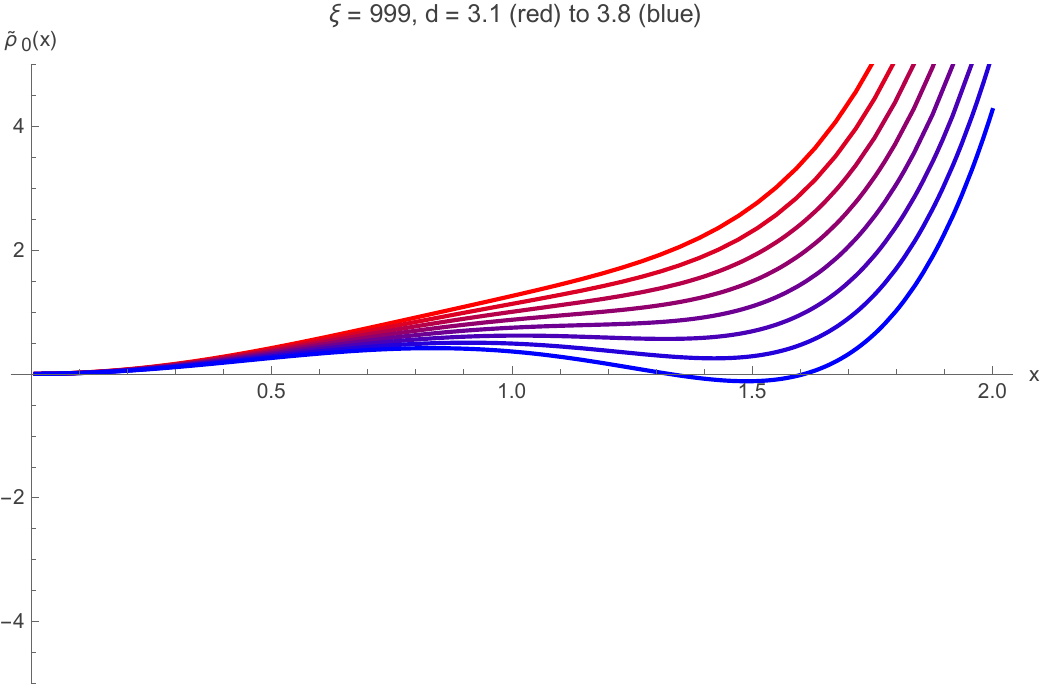}
    \end{minipage}
    \caption{The total scalar density obtained after adding overspinning states. The left graph corresponds to $\xi=9$, with $d$ going from $3.3$ (red) to $3.9$ (blue) in increments of $0.1$, and density of states starts to become negative for $d \gtrsim 3.86$. The right graph corresponds to $\xi=999$, with $d$ going from $3.1$ (red) to $3.8$ (blue) in increments of $0.1$, and the density develops a negative region for $d \gtrsim 3.77$.}
    \label{fig:overspin}
\end{figure}

Here we demonstrate an alternative overspinning state above the black hole threshold that is also able to cure the above negativities, without introducing new ones. Since the state is above the threshold, the spectral gap is preserved. The conformal weight of the state is chosen to be:  
\begin{align}
    \p{H,\bar{H}}=\p{\frac{17}{12},\frac{3}{4}}\xi
\end{align}
At leading semiclassical order, the bulk mass and spin are: 
\begin{align}
    \mathbf{M}\simeq\frac{1}{12}, \quad \mathbf{J}\simeq\pm\frac{1}{3}
\end{align}
Thus, the bulk geometry can be interpreted as an overspinning black hole with the above specified mass and angular momentum and no event horizon. We again add two seed spin-parity states each with a degeneracy of two. This yields the following density of states of the overspinning state: 
\begin{align}
    \rho^{\text{os}}_j=&\frac{2d}{\sqrt{t\bar{t}}}\sum_{s=1}^\infty \frac{1}{s}S(j,J;s) \cos\p{\frac{2\pi}{s}\sqrt{\frac{5}{3}}\sqrt{\xi \bar t}}\cosh \p{\frac{2\pi}{s}\sqrt{\xi t}}+ \notag \\&\frac{2d}{\sqrt{t\bar{t}}}\sum_{s=1}^\infty \frac{1}{s}S(j,-J;s) \cos\p{\frac{2\pi}{s}\sqrt{\frac{5}{3}}\sqrt{\xi  t}}\cosh \p{\frac{2\pi}{s}\sqrt{\xi \bar t}}
\end{align}
The negativities at the threshold and at the large-spin limit can be canceled by the mechanism described above. Positivity at finite spin, near extremality, and fixed $t$, follows directly from \cite{DiUbaldo:2023hkc}, since the leading exponential behavior is not changed but only the argument of the cosine function. The proof of the $j=0$ positivity for $\xi t \lesssim \mathcal{O}(1)$ follows straightforwardly from \cite{DiUbaldo:2023hkc}. We first establish the positivity of the sum of $s=1,2$ terms in the large $\xi$ limit:
\begin{align}
   \left. \frac{t}{2}\tilde{\rho}_0{}^{\text{total}}\right|_{s\leq 2}=2\sinh^2{x}+&2d\p{\cos\p{\sqrt{\frac{5}{3}}x}\cosh(x)-1}\notag \\
   +&d(-1)^J\p{\cos\p{\sqrt{\frac{5}{3}}\cdot\frac{x}{2}}\cosh\p{\frac{x}{2}}-1}
\end{align}
which will impose the following condition on $d$: 
\begin{align}
    &d \lesssim 2.195, \quad x_{\text{min}}\approx 1.161 \quad (J \hspace{1.5mm}\text{even}) \notag \\
    &d \lesssim 2.574, \quad x_{\text{min}}\approx 1.238 \quad (J \hspace{1.5mm}\text{odd})
\end{align}
For $s\geq3$, we write $y=x/s$ and define:
\begin{align}
F_s(y)=&
2\p{\phi(s)-\mu(s)}\sinh^2(2y)\notag\\
&+2d\,c_s(J)
\left[\cos\p{\sqrt{\frac53}\,y}\cosh y-1\right].
\end{align}
Each $s$-term contributes $F_s(x/s)/s$. Since $F_s(0)=F_s'(0)=0$, requiring $F_s''(y)>0$ for all $y\geq0$ is thus sufficient for positivity. Using 
\begin{align}
    \abs{c_s(J)}\leq\phi(s), \quad \abs{\mu(s)}\leq 1, \quad \phi(s\geq3)\geq2
\end{align}
the strongest inequality is given by: 
\begin{align}
    \frac{6 \cosh (4 x)}{\sqrt{15} \sinh (x)+\cosh (x)}>d
\end{align}
Minimizing the left-hand side yields the following bound:
\begin{align}
    d\lesssim 4.44, \quad x_{\text{min}}\approx 0.175
\end{align}
All of these bounds are compatible with $d=2$ in the original proposal. Moreover, the quantization of the spin $J=\frac{2}{3}\xi$ imposes a condition on the central charge $\frac{c-1}{36}\in \mathbb{Z}_+$. It is also useful to compare the above bounds with the whole sum, where by summing to $s=300$ we can check how the bound is affected by different values of the central charge $\xi$. For $\xi=9$, the density becomes negative for $d \gtrsim 3.86$, and for larger central charge $\xi=999$, the density develops a negative region for $d \gtrsim 3.77$, as shown in Fig. \ref{fig:overspin}. 

\begin{figure}[h]
    \centering
    \includegraphics[width=0.8\linewidth]{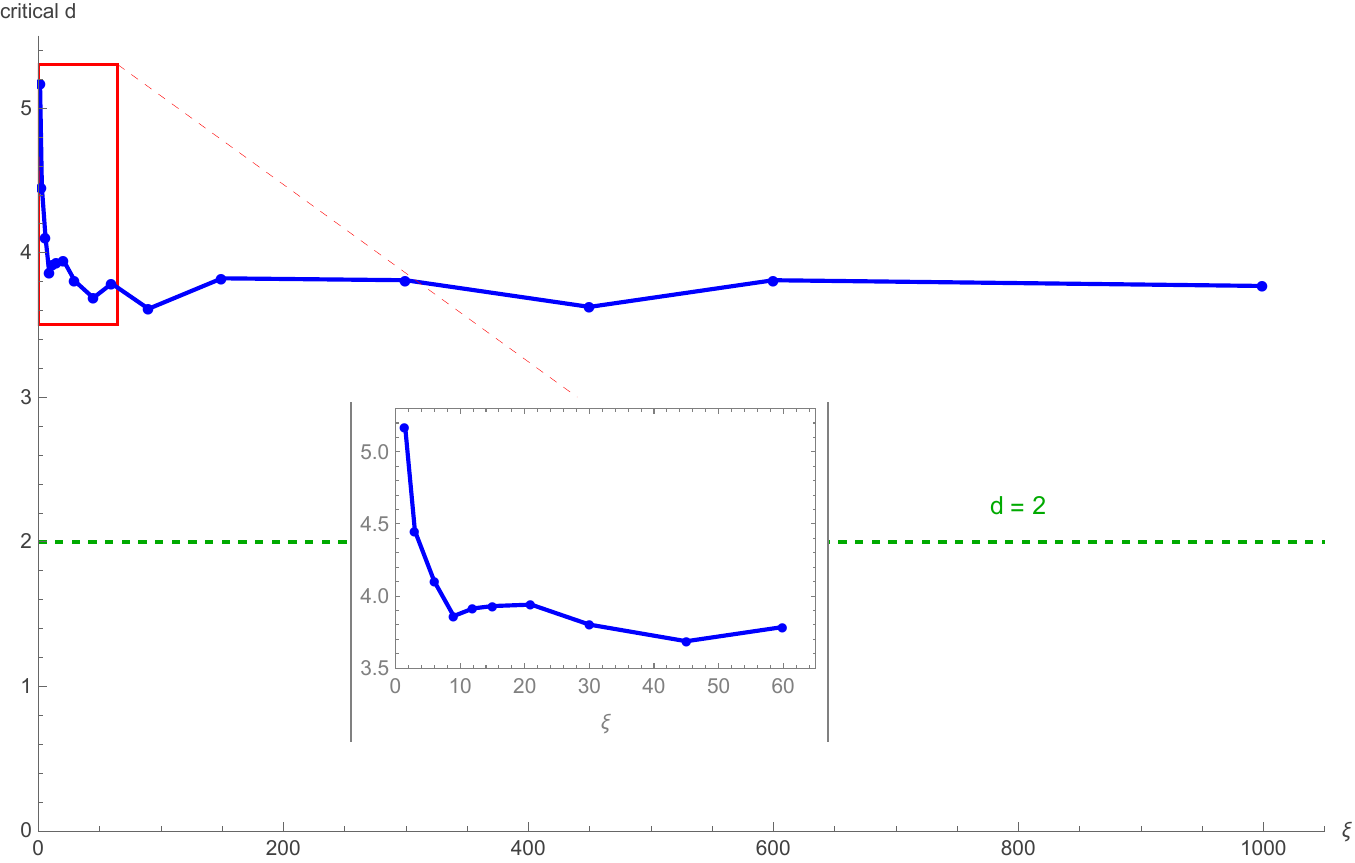}
    \caption{This graph shows how the bound on $d$ is affected by the central charge $\xi$, which ranges from $\xi=\frac{3}{2}$ to $\xi=999$. The sum is performed for $s=300$. For small values of $\xi$, the bound on $d$ initially dips and eventually stabilizes around $d \sim 3.8$.  }
    \label{fig:dvsxi}
\end{figure}

We can also gather more data points by plotting how $d$ is affected by $\xi$. For small values of the central charge, the bound on $d$ initially dips but eventually stabilizes well above $d=2$, as shown in Fig. \ref{fig:dvsxi}. 

\begin{figure}[t]
\centering
\begin{tikzpicture}[scale=1.0]

  \def\L{3.5}
  \def\s{4}

  \fill[black!8] (0,0) -- (-\L,\L) -- (\L,\L) -- cycle;
  \fill[black!20] (0,0) -- (-\L,-\L) -- (\L,-\L) -- cycle;
  \fill[black!14] (0,0) -- (\L,\L) -- (\L,-\L) -- cycle;
  \fill[black!14] (0,0) -- (-\L,\L) -- (-\L,-\L) -- cycle;

  \draw[semithick] (-\L,-\L) -- (\L,\L);
  \draw[semithick] (-\L,\L) -- (\L,-\L);

  \draw[semithick,dashed] (0,-1.5) -- (\L,{\L-1.5});
  \draw[semithick,dashed] (0,-1.5) -- (-\L,{\L-1.5});

  \draw[->,semithick] (-\L,0) -- (\L+0.3,0) node[right] {$\mathbf{J}$};
  \draw[->,semithick] (0,-\L) -- (0,\L+0.3) node[above] {$\mathbf{M}$};

  \fill (0,-1.5) circle (1.2pt);
  \node[left=3pt,font=\footnotesize] at (0,-1.5) {Empty AdS$_3$};


  \fill[black!30!orange] ({0.333*\s},{0.0833*\s}) circle (1.2pt);
  \fill[black!30!orange] ({-0.333*\s},{0.0833*\s}) circle (1.2pt);
  \draw[black!30!orange,thin,->] (\L+0.8,2.2) -- ({0.333*\s+0.06},{0.0833*\s+0.06});
  \draw[black!30!orange,thin,->] (\L+0.8,2.2) -- ({-0.333*\s+0.15},{0.0833*\s+0.06});
  \node[black!30!orange,font=\footnotesize,right] at (\L+0.8,2.2)
    {$\mathbf{M}\simeq\tfrac{1}{12},\;\mathbf{J}\simeq\pm\tfrac{1}{3}$};

  \fill[black!60!red] ({0.25*\s},0) circle (1.2pt);
  \fill[black!60!red] ({-0.25*\s},0) circle (1.2pt);
  \draw[black!60!red,thin,->] (-\L-0.8,1.0) -- ({-0.25*\s-0.06},0.06);
  \draw[black!60!red,thin,->] (-\L-0.8,1.0) -- ({0.25*\s-0.15},0.06);
  \node[black!60!red,font=\footnotesize,left] at (-\L-0.8,1.0)
    {$\mathbf{M}\simeq0,\;\mathbf{J}\simeq\pm\tfrac{1}{4}$};

  \fill[black!50!blue] ({0.125*\s},{-0.125*\s}) circle (1.2pt);
  \fill[black!50!blue] ({-0.125*\s},{-0.125*\s}) circle (1.2pt);
  \draw[black!50!blue,thin,->] (\L+0.8,-0.2) -- ({0.125*\s+0.06},{-0.125*\s+0.04});
  \draw[black!50!blue,thin,->] (\L+0.8,-0.2) -- ({-0.125*\s+0.1},{-0.125*\s+0.04});
  \node[black!50!blue,font=\footnotesize,right] at (\L+0.8,-0.2)
    {$\mathbf{M}\simeq-\tfrac{1}{8},\;\mathbf{J}\simeq\pm\tfrac{1}{8}$};

  \fill[black!40!teal] ({0.1*\s},{-0.15*\s}) circle (1.2pt);
  \fill[black!40!teal] ({-0.1*\s},{-0.15*\s}) circle (1.2pt);
  \draw[black!40!teal,thin,->] (-\L-0.8,-1.6) -- ({-0.1*\s-0.04},{-0.15*\s+0.04});
  \draw[black!40!teal,thin,->] (-\L-0.8,-1.6) -- ({0.1*\s-0.08},{-0.15*\s+0.04});
  \node[black!40!teal,font=\footnotesize,left] at (-\L-0.8,-1.6)
    {$\mathbf{M}\simeq-\tfrac{3}{20},\;\mathbf{J}\simeq\pm\tfrac{1}{10}$};

\end{tikzpicture}
\caption{Schematic phase diagram of AdS$_3$ quotient geometries in the $(\mathbf J,\mathbf M)$ plane. The solid lines $\mathbf M=\pm\mathbf J$ separate nonextremal BTZ black holes (top), overspinning geometries (sides), and the defect/excess region (bottom). The dashed lines starting from empty AdS$_3$ denote the unitarity bound. The colored points are the examples discussed in the text, with their leading semiclassical charges shown in the labels; their plotted locations are schematic. At this order they satisfy $\mathbf M=\abs{\mathbf J}-\frac14$, since $T=-\xi/4$.}
\label{fig:phase-diagram}
\end{figure}
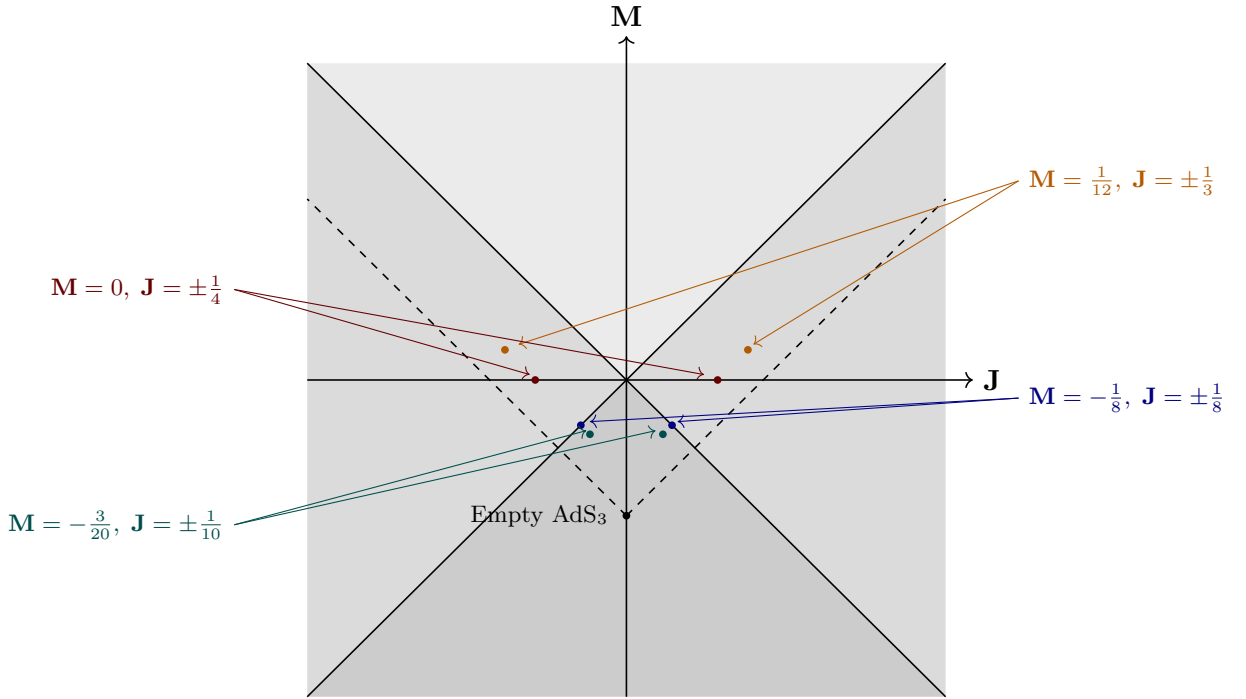

\section{Bulk geometries of the spinning states}
\label{section3}
The bulk interpretation of these spinning states is not unique, since their conformal weights only determine the asymptotic mass and angular momentum but do not specify the interior bulk completion. The full CFT$_2$, however, contains further microscopic information. Motivated by a minimal matter content perspective, we interpret the spinning states below the black hole threshold as spinning defects, which are supported by a delta-function source, and the overspinning state above the BTZ threshold as overspinning BTZ geometries, which are smooth Lorentzian geometries free of singularities. It is also instructive to visualize these spinning states in the $\mathbf{M}, \mathbf{J}$ phase diagram, shown in Fig. \ref{fig:phase-diagram}. Both the spinning defects and overspinning BTZ geometries contain causal pathologies in their Lorentzian geometries. Interpreting these overspinning states as overspinning BTZ geometries has the advantage of preserving the metric degrees of freedom in the gravitational path integral, but the Lorentzian geometry alone does not establish their contribution to the Euclidean gravitational path integral. Their quasi-Euclidean continuation, despite being smooth, fails the Kontsevich–Segal–Witten criterion \cite{Basile:2023ycy}. We return to this obstruction later in this section.

In this section we study several classical and semiclassical aspects of these spinning geometries \cite{Miskovic:2009uz,Martinez:2019nor,Briceno:2021dpi,Li:2024rma}, including the computation of scalar propagators on these backgrounds. For the extremal defect, one can use the standard method of quantizing scalar fields on a curved background, while for overspinning geometries we propose a quotient motivated image expression whose leading contribution reproduces the semiclassical heavy-light Virasoro vacuum block. On the other hand, the causally well-behaved interior completions such as spinning strings or rotating shells can remove the CTC region, as in the interpretation adopted by \cite{DiUbaldo:2023hkc}. They define different fluctuation problems: some nonminimal geodesics probe the removed region \cite{Caminiti:2025gyv}, and the scalar modes must satisfy additional boundary or matching conditions. Thus, the complete scalar correlator is not simply defined by the asymptotic charges alone. 

\subsection{Sub-extremal spinning defect}
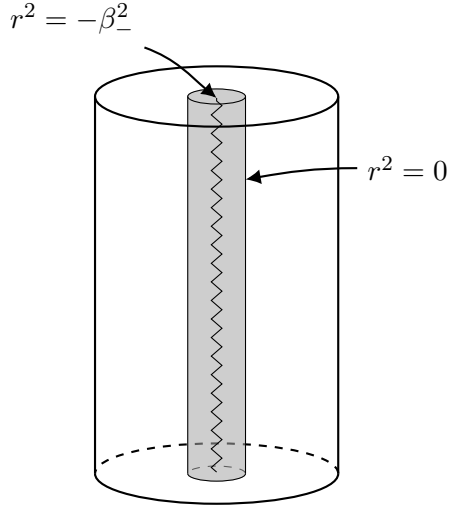
\begin{figure}[t]
\centering
\begin{tikzpicture}
\draw[dashed, thick] (1.6,0) arc(0:180:1.6 and .4);

\draw[dashed] (.38,0) arc(0:180:.38 and .1);
\begin{scope}[opacity=.55,transparency group]
    \filldraw[black!35] (0,0) ellipse (.38 and .1);
    \filldraw[black!35] (0,5) ellipse (.38 and .1);
    \filldraw[black!35] (-.38,0) rectangle (.38,5);
\end{scope}
\draw (0,5) ellipse (.38 and .1);
\draw (-.38,0) arc(180:360:.38 and .1);
\draw (-.38,0) -- (-.38,5);
\draw (.38,0) -- (.38,5);

\draw[
    thin,
    decorate,
    decoration={zigzag, segment length=7pt, amplitude=2pt}
] (0,0.02) -- (0,4.98);

\draw[thick] (0,5) ellipse (1.6 and .4);

\draw[thick] (-1.6,0) -- (-1.6,5);
\draw[thick] (1.6,0) -- (1.6,5);

\draw[thick] (-1.6,0) arc(180:360:1.6 and .4);

\draw[-{Latex[length=2mm,width=2mm]}, thick] 
    (1.85,4.05) to[out=180,in=10] (.38,3.9);
\node[right] at (1.85,4.05) {$r^2=0$};

\draw[-{Latex[length=2mm,width=2mm]}, thick] 
    (-0.95,5.65) to[out=-20,in=140] (0,4.98);
\node[above left] at (-0.95,5.65) {$r^2=-\beta_-^2$};

\end{tikzpicture}
\caption{A sub-extremal spinning defect depicted in global coordinates ($t$ runs up and $r$ goes radially outward). The conical defect is located at \(r^2=-\beta_-^2\) and is enclosed by a shaded tubular region containing closed timelike curves, which remains accessible to an outside observer. The boundary of this region is at \(r^2=0\), beyond which the spacetime is causally well behaved. }
\label{fig:spinningdefect}
\end{figure}

The sub-extremal defect is given by the BTZ metric\footnote{We set $l=1$ from now on.}:
\begin{align}
\label{metricspinning}
    \dd s^2=-\p{r^2+\abs{\mathbf{M}}}\dd t^2+ \p{r^2+\abs{\mathbf{M}}+\frac{\mathbf{J}^2}{4r^2}}^{-1}\dd r^2+r^2 \dd \phi^2 -\mathbf{J}\dd \phi \hspace{1mm} \dd t
\end{align}
This can be viewed as the negative mass spectrum of the BTZ black hole. The above metric can be obtained from the identification of empty global AdS$_3$:  
\begin{align}
\label{emptyads}
\dd s^2=-\p{1+\hat{r}^2}\dd \hat{t}^2+\p{1+\hat{r}^2}^{-1} \dd \hat{r}^2+\hat{r}^2
\dd \hat{\phi}^2 
\end{align}
with the transformation:
\begin{align}
   \hat{r}^2=\frac{r^2+\beta_{-}^2}{\beta_+^2-\beta_-^2},\quad      \hat{t}=\beta_-\phi+\beta_+t, \quad \hat{\phi}=\beta_+\phi +\beta_- t
\end{align}
which imposes the identification $(\hat{t},\hat{\phi})\sim (\hat{t}+2\pi\beta_-,\hat{\phi}+2\pi \beta_+)$, where we have\footnote{In \cite{Li:2024rma} we assumed $\mathbf{J}<0$, which can be obtained by $t\to-t$.}:
\begin{align}
   -\mathbf{M}=\beta_+^2+\beta_-^2,\qquad \mathbf{J}=2\beta_+\beta_-
\end{align}
In the spinning defect completion adopted here, the fixed loci of the rotations in the spacelike and timelike planes are treated as conical singularities with angular deficits. It should be noted  that the full helical quotient is actually fixed point free, but retaining it without a matter source makes the bulk interpretation rather subtle. Since the locus at $\hat{r}^2=0$ provides a natural place for a conical matter source, we will adopt such an interior completion. In the quotient spacetime, the conical defect at $\hat r^2=0$, or equivalently at $r^2=-\beta_-^2$ is surrounded by a tubular region $-\beta_-^2\leq r^2<0$ containing CTCs, which is accessible to an outside observer. This is depicted in Figure \ref{fig:spinningdefect}. The other defect of $\hat t$ is located at $r^2=-\beta_+^2$, and does not lie on the real AdS$_3$ hyperboloid. The geometry can now be seen as arising from a matter source supported by the stress-energy tensor at $\hat{r}^2=0$: 
\begin{align}
    \sqrt{g^{(2)}}T_{\mu \nu} u^\mu u^\nu=\frac{1-\beta_+}{4G}\delta^2(\hat{r})
\end{align}
where $u^\mu$ is tangent to the defect worldline and $g^{(2)}$ is the determinant of the metric transverse to it. This equation records the localized energy density and the spin is encoded in the helical holonomy around the source. 
From the embedding space $\mathbb{R}^{2,2}$\footnote{We adopt the convention where $\dd s^2 = -(\dd X^0)^2 - (\dd X^1)^2 + (\dd X^2)^2 + (\dd X^3)^2$.}, the spinning defect geometry can be obtained by quotienting along the orbits of the vector field \cite{Miskovic:2009uz}: 
\begin{align}
    \Theta=\beta_+J_{23}+\beta_-J_{01}
\end{align}
where $J_{ab}=2X_{[a}\partial_{b]}$ and the embedding coordinates are given by: 
\begin{align}
\label{subextremalembedding}
    &X^0=\sqrt{\alpha(r)+1}\cos\p{\phi_{01}},
    \hspace{1.8cm}
    X^2=\sqrt{\alpha(r)}\cos\p{\phi_{23}} \notag \\*
    &X^1=\sqrt{\alpha(r)+1}\sin\p{\phi_{01}},
    \hspace{1.84cm}
    X^3=\sqrt{\alpha(r)}\sin\p{\phi_{23}}
\end{align}
together with
\begin{align}
    \alpha(r)=\frac{r^2+\beta_-^2}{\beta_+^2-\beta_-^2}
\end{align}
Note that unlike the static defect which only involves a rotation in the spatial generator, the spinning defect is a combination of rotations in the $(01)$-timelike and $(23)$-spatial planes with angles $\phi_{23}=\beta_+\phi+\beta_-t$ and $\phi_{01}=\beta_+t+\beta_- \phi$. The two loci arise from the separate $J_{23}$, $J_{01}$ generators, located at $\a(r)=0$ and $\a(r)=-1$ respectively. The combined full helical action is nevertheless fixed point free. The scalar correlator was studied in \cite{Li:2024rma}. For this calculation, we use the radial coordinate:
\begin{align}
    z=\frac{r^2+\beta_-^2}{r^2+\beta_+^2}
\end{align}
with regularity imposed at the source endpoint $z = 0$ and the normalizable falloff at $z = 1$. When the region containing CTCs is included, the resulting correlator is given by the following mode sum\footnote{In the original radial coordinate $r$, the boundary lies at $r=\infty$, and we use $r^{\Delta}$ to extrapolate the normalizable mode. In \cite{Li:2024rma}, the extrapolate dictionary was used without the factor $(\beta_+^2-\beta_-^2)^{\Delta/2}$. To match the bulk result exactly and take the $\beta_-\to\beta_+$ extremal limit, it is useful to include this factor: $\psi^-_{k\ell}(\Delta)=\lim_{r\to \infty}r^\Delta R_{+}\left[z(r)\right]=\lim_{z\to1}\p{\frac{z\beta_+^2-\beta_-^2}{1-z}}^{\Delta/2}R_+(z)$.}:
\begin{align}
\label{subextremalpropagator}
    G^+(t,\phi)&=\frac{\p{\beta_+^2-\beta_-^2}^{\Delta}}{\beta_+-\beta_-}\sum_{\ell=-\infty}^{\infty} \sum_{k =0}^{\infty} \frac{ \Gamma\p{\Delta+k+2\alpha_{k\ell}}\Gamma\p{\Delta+k}}{\Gamma\p{\Delta}^2\Gamma\p{1+k+2\alpha_{k\ell}}\Gamma\p{1+k}}e^{-i\omega_{k\ell} t} e^{i\ell\phi}, \notag \\
   \alpha_{k\ell}&=\frac{\ell \beta_+ +\omega_{k\ell} \beta_-}{2(\beta_+^2-\beta_-^2)}, \qquad  \omega_{k\ell}=\ell+(2k+\Delta)\p{\beta_++\beta_-}
\end{align}
The above correlator should be viewed as containing the branch with $\a_{k\ell}>0$. The other regular branch $z^{-\a}$ with $\a<0$ gives the modes $\omega_{k\ell}=-\ell+(2k+\Delta)(\beta_+-\beta_-)$. The Feynman propagator includes both branches. Moreover, the full-range radial integral for the scalar field normalization constant includes the CTC region and should be viewed as an analytic prescription for the Klein–Gordon inner product, not a positive norm on a global Cauchy surface. The associated image expression is given by
\begin{align}
\label{moisubex}
    G \p{t,\phi}=\sum_{m\in \mathbb{Z}} \bigg[ \frac{4}{\beta_+^2-\beta_-^2}\sin\p{\frac{\beta_+-\beta_-}{2}\p{\phi+2\pi m-t}}\sin\p{\frac{\beta_++\beta_-}{2}\p{\phi+2\pi m+t}} \bigg]^{-\Delta}
\end{align}
Note that for generic parameters the terms do not decay as $\abs{m}\to \infty$, and the image sum diverges. It can be shown that using the appropriate regularization procedure proposed in \cite{Li:2024rma}, the above infinite image sum is equivalent to the bulk mode sum (\ref{subextremalpropagator}). From the dual CFT$_2$ point of view, the leading image reproduces the heavy-light Virasoro vacuum block. To construct a crossing-symmetric correlator, one then includes the vacuum blocks across all channels \cite{Maloney:2016kee,Anous:2017tza,Hijano:2019qmi}, which in the bulk corresponds to the angular images.  Thus, within the vacuum-block approximation, the bulk and boundary correlators are controlled by the same Virasoro vacuum block, making their universality manifest.

However, if we choose to excise the $r^2<0$ region or replace the interior geometry by introducing certain matter fields such as the spinning string or matter shells, the radial problem is qualitatively different. At a finite cutoff $z=z_c$, with $\frac{\beta_-^2}{\beta_+^2}\leq z_c< 1$, both independent hypergeometric solutions $R_{\pm}(z)$ are regular. Regularity therefore no longer selects a unique radial solution; instead we can in general take certain linear combinations of $R(z)_\pm$ depending on the boundary or junction condition at the cap. This modifies the mode spectrum, the Klein-Gordon normalization, and the resulting bulk propagator. Since the leading semiclassical contribution of the $m=0$ geodesic does not probe the cap (or the string worldsheet), it may continue to reproduce the $t-$channel semiclassical vacuum block. However, nonminimal geodesics that probe the cap can instead be modified or removed \cite{Caminiti:2025gyv}. The full image sum and its interpretation as a sum over modular images of the vacuum block, no longer follow from the quotient geometry alone. From the CFT$_2$ perspective, this corresponds to additional data beyond the universal vacuum Virasoro block. The same can be said about the rotating shell: Israel junction conditions determine the gluing of the background geometries, and scalar fields require additional matching conditions at the shell. Consequently, different matter completions with the same asymptotic charges $(\mathbf{M},\mathbf{J})$ can lead to different scalar correlators.

\subsection{Overspinning geometries}
The overspinning geometries belong to the BTZ family with $\abs{\mathbf{M}}<\abs{\mathbf{J}}$. When the mass parameter is positive, this can be viewed as an overspinning BTZ black hole. Unlike higher-dimensional black holes, the geometry is completely smooth since the spinning BTZ black hole does not have a curvature singularity, or any physical singularity, and the spacetime is smooth across $r^2=0$. However, the absence of any real event horizon exposes the region $r^2<0$, exhibiting causal pathologies for the spacetime. To see what happens at $r^2=0$, set $x=r^2$. The metric becomes
\begin{align}
\label{overspin-xmetric}
    \dd s^2&=-(x-\mathbf M)\dd t^2-\mathbf J\dd t\dd\phi+x\dd\phi^2
    +\frac{\dd x^2}{D(x)},\notag\\
    D(x)&=4\left(x-\frac{\mathbf M}{2}\right)^2+\mathbf J^2-\mathbf M^2>0.
\end{align}
Since $\det g=-1/4$, the metric and its inverse are regular for all real $x$, including $x=0$. In fact, the angular Killing vector $K=\partial_\phi$ obeys $K^2=x$ and $K\cdot\partial_t=-\mathbf J/2$, so it becomes null but does not contract to a point at $x=0$. The radial position $x=0$ is in fact a timelike hypersurface (cylinder) with a nondegenerate Lorentzian induced metric in the rotating geometries. Note that for $x>0$, $t$ is the time function, while for $x<0$ the geometry contains CTCs due to $\dd \phi^2$ becoming timelike. 

This gives two different completion choices. Treating $x=0$ as its end point requires a timelike boundary and specifying the boundary condition. Extending to all ranges of $x$ will instead give the maximal-quotient together with a second asymptotic end in Lorentzian signature with CTCs. Retaining the full quotient has the advantage of being source free and the scalar correlator can be expressed using the method of images. However, the second asymptotic end requires a boundary prescription and identifying its contribution to the one-torus Euclidean path integral remains open.

In terms of the event horizons $r_\pm$, the metric becomes complex, and we need to find a new combination of the mass and spin parameters that is real. This can be achieved by defining: 
\begin{align}
    a=\frac{\sqrt{\abs{\mathbf{J}}+\mathbf{M}}}{2}, \quad  b=\frac{\sqrt{\abs{\mathbf{J}}-\mathbf{M}}}{2}
\end{align}
The new combination is now given by: 
\begin{align}
    \gamma_+=a+b, \quad \gamma_-=a-b
\end{align}
where the mass $\mathbf{M}$ and the angular momentum $\mathbf{J}$ are given by: 
\begin{align}
    2\g_+ \g_-=\mathbf{M}, \quad \g_+^2+\g_-^2=2(b^2+a^2)=\abs{\mathbf{J}}
\end{align}
Similar to the sub-extremal defect, the overspinning geometry can also be obtained from identifications of the embedding space $\mathbb{R}^{(2,2)}$, but with the following combination: 
\begin{align}
    \Theta=b\p{J_{01}+J_{23}}-a\p{J_{03}+J_{12}}
\end{align}
Note that this is no longer a simple rotation or boost in different planes, but rather a combination of both involving separate planes. As a result, the embedding space coordinates are more complicated compared to the sub-extremal defect. Writing $C_\pm(y)=\cos y\pm\sin y$, they are given by \cite{Briceno:2021dpi}: 
\begin{align}
X^0 &= \tfrac12\sqrt{A(r)+1}\,\cosh(au)\,C_-(bv)
     + \tfrac12 S(r)\,\sinh(au)\,C_+(bv), \notag\\
X^1 &= \tfrac12\sqrt{A(r)+1}\,\cosh(au)\,C_+(bv)
     - \tfrac12 S(r)\,\sinh(au)\,C_-(bv), \notag\\
X^2 &= \tfrac12\sqrt{A(r)+1}\,\sinh(au)\,C_+(bv)
     - \tfrac12 S(r)\,\cosh(au)\,C_-(bv), \notag\\
X^3 &= \tfrac12\sqrt{A(r)+1}\,\sinh(au)\,C_-(bv)
     + \tfrac12 S(r)\,\cosh(au)\,C_+(bv).
\label{overspinningembedding}
\end{align}
The coordinates $u,v$ are light-cone coordinates defined as $u=t-\phi$ and $v=t+\phi$, and 
\begin{align}
A(r) &= \frac{1}{2ab}
        \sqrt{(a^2+b^2+r^2)^2-4a^2r^2}, \quad S(r)  = \frac{a^2-b^2-r^2}{2ab\sqrt{A(r)+1}}.
\end{align}
where $S(r)^2=A(r)-1$ and selects the signed square root that yields the correct metric. The above quotient does not introduce any fixed points, since the rotation and boost in $\Theta$ act together. Although $A(r)=1$ at $r^2=a^2-b^2$, which makes the terms proportional to $\sqrt{A-1}$ vanish in the embedding space coordinates, the full quotient generator does not vanish there. In fact, the only fixed point of the embedding space lies in the origin of $\mathbb R^{2,2}$, which lies outside the AdS$_3$ pseudosphere \cite{Hulik:2019pwr}. Together with the regularity of the metric, this shows that the maximal quotient has no conical or orbifold singularity. 

The appearance of hyperbolic functions in the generator hints that the spacetime is no longer obtained by identification of purely elliptic generators, but a combination of elliptic and hyperbolic generators, which makes the geometry contain a formally chiral temperature \cite{Hulik:2019pwr}. Recall that it is useful to define a left/right-moving temperature for the BTZ black hole: 
\begin{align}
    T_L=\frac{r_+-r_-}{2\pi}=\frac{\sqrt{\mathbf{M}-\mathbf{J}}}{2\pi}, \qquad T_R=\frac{r_++r_-}{2\pi}=\frac{\sqrt{\mathbf{M}+\mathbf{J}}}{2\pi}
\end{align}
For the range $\mathbf J>\abs{\mathbf M}$ discussed above, 
\begin{align}
    T_L=\frac{i\sqrt{\mathbf J-\mathbf M}}{2\pi},
    \qquad
    T_R=\frac{\sqrt{\mathbf J+\mathbf M}}{2\pi}
\end{align}
with one imaginary and one real parameter. Since the overspinning geometries has neither an event horizon nor a Euclidean regularity condition that fixes the thermal circle, these quantities are not Hawking temperatures. We only use the phrase \textit{partially thermal} in this formal holonomy sense. This chiral parameter can also be related to the modular variable in the dual CFT$_2$ character given by: 
\begin{align}
\label{overspinpart}
   Z_{\text{overspin}}(q,\bar q)=\frac{
q^{\,H-\xi}\bar q^{\,\bar H-\xi}}{\eta(\tau)\eta\p{-\bar{\tau}}}, \qquad
q=e^{2\pi i\tau},\quad \bar q=e^{-2\pi i\bar\tau},
\end{align}
Reproducing the above expression is another motivation for retaining the full-quotient, since introducing different matter completions will in general change the fluctuation problem. Cutting the geometry at $x=0$ changes the domain of fluctuation operators and requires an inner boundary condition, while replacing the interior by a string or matter shell introduces junction conditions and possible matter mode. For a general matter completion, there is no reason to expect the final answer to factorize as $Z_{\mathrm{full}}=Z_{\mathrm{overspin}}Z_{\mathrm{matter}}$ and even if it holds, we still need $Z_{\mathrm{matter}}=1$ for an exact match, which requires a rather non-trivial cancellation for the matter determinant. 

The analysis of a scalar field in the overspinning geometry is rather different from the sub-extremal case, since the arguments of the hypergeometric functions will now involve complex arguments. Before proceeding to the discussion of the mode problem, it is instructive to compute the leading geodesic:
\begin{align}
    \mathcal{L}=\cosh^{-1}\p{X_0X'_0+X_1X'_1-X_2X'_2-X_3X'_3}
\end{align}
By evaluating $X_A'$ at $t=\phi=0$, and taking the asymptotic limit $r\to \infty$, one immediately finds, using (\ref{overspinningembedding}): 
\begin{align}
    \mathcal{L}=\log\left[ \frac{1}{ab}\sin\bigg(b\p{t+\phi}\bigg)\sinh\bigg(a(\phi-t)\bigg)  \right]
\end{align}
Following the sub-extremal case, the periodicity in $\phi$ can be recovered by summing over all image geodesics, which motivates the following form of the correlator: 
\begin{align}
\label{moii1}
    G\p{t,\phi}=\sum_{m\in \mathbb{Z}} \bigg[ \frac{1}{ab}\sin\bigg(b\p{t+\phi+2\pi m}\bigg)\sinh\bigg(a(\phi-t+2\pi m)\bigg) \bigg]^{-\Delta}
\end{align}
The leading geodesic can be viewed as the dual CFT$_2$ vacuum block continued to the range $\abs{\mathbf{J}}>\abs{\mathbf{M}}$, and does not probe into the region $x<0$, but the rest of the image geodesics can in fact enter the region containing CTCs. Removing this region will indeed change how we periodize the vacuum block to obtain a crossing symmetric correlator. 

The obstruction to a first-principles derivation of (\ref{moii1}) is already visible in the radial equation. For what follows, we take $\mathbf{M}>0$ and use the combination $\g_\pm$ to write the metric as: 
\begin{align}
    \dd s^2=-\p{r^2-2\g_+\g_-}\dd t^2-(\g_+^2+\g_-^2)\dd t \dd \phi+\p{r^2-2\g_+\g_-+\frac{(\g_+^2+\g_-^2)^2}{4r^2}}^{-1}\dd r^2+r^2\dd \phi^2
\end{align}
Despite being able to put the metric in a real form, the structure of the Klein-Gordon equation remains unchanged. In the rotating black hole regime, the regular singularities of the differential equations are located at the event horizons $r^2_\pm$ and the conformal boundary. In order to put the solution into the hypergeometric form, we need to map these singularities to the ones of the hypergeometric equation, which are now located at complex radii: 
\begin{align}
    r_\pm=a\pm i b
\end{align}
with the following coordinate transformation: 
\begin{align}
    z=\frac{r^2-r_+^2}{r^2-r_-^2}
\end{align}
the radial solution of the scalar field $\Phi=e^{-i\omega t}e^{i\ell \phi}R(z)$ can be put into the hypergeometric form: 
\begin{align}
    R(z)=z^\a(1-z)&^{\D/2} {}_2F_1\left[\frac{\Delta}{2}-\frac{i\p{\omega+\ell}}{4a},\frac{\D}{2}-\frac{\omega-\ell}{4b},1+2\a;z\right],\notag \\
    \a&=-\frac{\omega-\ell}{8b}-\frac{i\p{\omega+\ell}}{8a}.
\end{align}
where the other solution can be computed with $\a\to-\a$. In order to compute the scalar correlator, we need to single out one of the solutions. Since there is no defect or event horizon, we do not have a consistent way to impose boundary conditions. The natural boundary condition we imposed at $z=0$ for the sub-extremal defect is now at a complex radial point, whose physical significance is not clear. If we formally impose the boundary condition at $z=0$, the fall-off boundary condition at infinity yields the following poles of the gamma function: 
\begin{align}
    \omega_L=\ell+2b\p{2k+\D},\qquad \omega_R=-\ell-2ai(2k+\D)
\end{align}
where the left-moving frequency is real and quantized like normal modes, while the right-moving frequency yields quasinormal modes with an imaginary part. To establish the complex frequencies as actual quasi-normal modes of the overspinning geometry requires a justification or physical motivation for the boundary condition. Moreover, a simple $i\epsilon$ prescription shows that the expansion of (\ref{moii1}) only contains real frequency-modes and $\omega_R$ are not poles of the propagator (see appendix \ref{overspinpoles}). Thus, we will only treat the complex modes as a formal pole structure resulting from the chosen boundary condition, instead of the actual quasi-normal modes of the geometry.

\subsection{Extremal spinning defect}
The extremal spinning defect is obtained from an elliptic--parabolic quotient. In the limit $\abs{\mathbf{M}}=\abs{ \mathbf{J}}$ of (\ref{metricspinning}), the metric takes the form: 
\begin{align}
\label{metricextremal}
    \dd s^2=-\p{r^2+2\beta_+^2}\dd t^2+ \frac{r^2 \dd r^2}{\p{r^2+\beta_+^2}^2}+r^2 \dd \phi^2 -2\beta_+^2\dd \phi \hspace{1mm} \dd t
\end{align}
where $\beta_+=\sqrt{\abs{\mathbf{M}}/2}$. We see that the locations of the two conical defects merge in the extremal limit into a single one located at $r^2=-\beta_+^2$, which is again surrounded by a tube of CTCs. This also changes the solutions of the scalar wave equation. Instead of being hypergeometric functions, the solutions are now given by the confluent hypergeometric functions with two singularities. This can be viewed as a limiting case for the hypergeometric function, whose two regular singularities merge into one irregular singularity. The scalar wave equation can then be cast as the confluent hypergeometric equation with the following transformation: 
\begin{align}
    z=\frac{\beta_+^2}{r^2+\beta^2_+}
\end{align}
Using $\Phi\p{\textbf{x}}=e^{-i\omega t}e^{i\ell \phi}R(z)$, the wave equation is then: 
\begin{align}
    \frac{d^2 R(z)}{dz^2}-R(z)\p{\frac{\p{z-1}\omega^2}{4z \beta_+^2}+\frac{\ell \omega}{2\beta_+^2}+\frac{\p{1+z}\ell^2}{4z \beta_+^2}+\frac{\Delta\p{\Delta-2}}{4z^2}}=0
\end{align}
The solution is given by the Whittaker function (confluent hypergeometric function), and has two singularities at $z=0$ and $z=\infty$. The irregular singularity is located at $z=\infty$, representing the location of the defect. In this case, we have two linearly independent solutions: 
\begin{align}
    R_+(z) =M_{\frac{\omega -\ell}{4 \beta _+},\frac{\Delta -1}{2}}\left(\frac{z (\ell+\omega )}{\beta _+}\right), \qquad R_-(z)=W_{\frac{\omega -\ell}{4 \beta _+},\frac{\Delta -1}{2}}\left(\frac{z (\ell+\omega )}{\beta _+}\right)
\end{align}
As before, we need to impose the boundary condition at the location of the defect to single out one of the solutions. Let's define  $Q=\omega+\ell$. For the branch $Q>0$, $R_-(z)$ is regular and  the quantization condition can be computed using the asymptotic expansion near $z=0$, which gives: 
\begin{align}
    \frac{\ell-\omega +2 \Delta  \beta _+}{4 \beta _+}=-k, \quad k\in \mathbb{Z}_{\geq0}
\end{align}
This implies: 
\begin{align}
    \omega_{k\ell}=\ell+2\beta_+\p{2k+\Delta}, \quad k\in \mathbb{Z}_{\geq0}
\end{align}
Note that condition $\omega+\ell>0$ is satisfied. For these modes, we can define $Q=2p_{k\ell}$ where 
\begin{align}
    p_{k\ell}=\ell+\beta_+\p{2k+\Delta}
\end{align}
so the branch condition is equivalent to $p_{k\ell}>0$. The $Q<0$ branch then selects the regular solution given by:
\begin{align}
    W_{-\frac{\omega-\ell}{4\beta_+},\frac{\Delta-1}{2}}
    \left(-\frac{z(\omega+\ell)}{\beta_+}\right),
\end{align}
which can be obtained from the $Q>0$ branch by $(\omega,\ell)\to(-\omega,-\ell)$. This solution supplies the negative frequency modes $\omega^{(-)}_{k\ell}=\ell-2\beta_+(2k+\Delta)$, and both branches will contribute to the Feynman propagator, analogous to the case of the static defect, where both solutions with $\ell>0$ and $\ell<0$ combine into a result depending on $\abs{\ell}$ \cite{Berenstein:2022ico}.

As before, this can be plugged back into the solutions and the Klein-Gordon inner product can then be computed. The integration range is again from the location of the defect to the conformal boundary, so that the region containing CTCs is included. We leave the details to appendix \ref{extremalcorrelator} and quote the result here. For the $Q>0$ branch, we have: 
\begin{align*}
     \langle \Phi,\, \Phi \rangle=\mathcal{N}_{k\ell}^2=p_{k\ell}\G\p{1+k}\G\p{k+\Delta}
\end{align*}
Using the extrapolate dictionary together with the normalization constant, we can assemble the positive-frequency Wightman function: 
\begin{align}
\label{extremalpropagator}
    G^+\p{t,\phi}=\sum_{k=0}^{\infty}\sum_{\substack{\ell\in\mathbb Z\\p_{k\ell}>0}}
    \frac{\p{2\beta_+}^\Delta p_{k\ell}^{\Delta-1}\Gamma\p{\Delta+k}}
    {\Gamma\p{\Delta}^2\Gamma\p{1+k}}
    e^{-i\omega_{k\ell}t}e^{i\ell \phi},
\end{align}
The case for $Q=0$ is excluded from the analysis, since it does not admit a non-zero solution satisfying the boundary conditions for $\Delta>1$.  The negative-frequency function is simply given by:
\begin{align*}
    G^-(t,\phi)=G^+(-t,-\phi),
\end{align*}
and the Feynman propagator can be assembled using these. As a consistency check, this result can actually be obtained as the $\beta_+\to \beta_-$ limit of the positive frequency sub-extremal propagator (\ref{subextremalpropagator}). In the limit $\beta_+\to \beta_-$, we have: 
\begin{align}
    \alpha=\frac{\p{\omega+\ell}\beta_+}{2 y},\quad y=\beta_+^2-\beta_-^2 \to0
\end{align}
Using the standard asymptotics of the gamma functions: 
\begin{align}
    \frac{\Gamma\p{\Delta+k+2\alpha}}{\Gamma\p{1+k+2\alpha}} \sim \left(\frac{(\omega_{k\ell}+\ell)\beta_+}{y}\right)^{\Delta-1}
\end{align}
The quantization condition that does not vanish gives: 
\begin{align}
    \omega_{k\ell}=\ell+2\beta_+\p{2k+\Delta}
\end{align}
and combining with the prefactor, we have: 
\begin{align}
    \frac{y^\Delta}{\beta_+-\beta_-}
    \p{\frac{(\omega_{k\ell}+\ell)\beta_+}{y}}^{\Delta-1}
    \longrightarrow(2\beta_+)^\Delta p_{k\ell}^{\Delta-1}
\end{align}
This yields exactly the propagator for the extremal spinning defect (\ref{extremalpropagator}). Note that taking the same limit using the geodesic length, the final result is as expected: 
\begin{align}
\label{extremalpropa}
    G^+_{\mathrm{img}}(t,\phi)=\lim_{\epsilon\to0^+}\sum_{m\in\mathbb Z}
    \beta_+^\Delta(\phi-t+2\pi m+i\epsilon)^{-\Delta}
    \left[\sin\p{\beta_+(\phi+t+2\pi m-i\epsilon)}\right]^{-\Delta}
\end{align}
where the leading $m=0$ geodesic is again the extremal limit of the heavy-light vacuum block below the black hole threshold, while the other contributions are the vacuum blocks in all other channels. We have also introduced the $i\epsilon$ prescription for image geodesics, where by expanding the sine factor and applying Poisson resummation, one can reproduce (\ref{extremalpropagator}) including the condition $p_{k\ell}>0$ and the normalization, as shown in appendix (\ref{appenb1}).

\subsection{Euclidean Geometry and KSW Criterion}
Most of our bulk discussion has focused on the properties of the Lorentzian geometries, and whether these are admissible Euclidean saddles remains to be established. To preserve the Lorentzian charges, the Wick rotation $t=-i\tau$ yields quasi-Euclidean saddles whose metrics are complex. However, these quasi-Euclidean saddles do not satisfy the Kontsevich–Segal–Witten criterion \cite{Kontsevich:2021dmb,Witten:2021nzp}, as shown in \cite{Basile:2023ycy} for the overspinning geometries. The motivation behind the KSW criterion is to identify whether complex metrics can support fluctuations of quantum fields, and the condition reads: 
\begin{align}
    \mathrm{Re}\left[\rho g^{\mu_1\nu_1}\cdots g^{\mu_q\nu_q}F_{\mu_1 \cdots \mu_q} F_{\nu_1\cdots \nu_q}\right]>0, \quad \rho=\sqrt{\mathrm{det}g_{\mu \nu}}. 
\end{align}
In the real coordinate basis $(\tau,\phi,x)$, the quasi-Euclidean BTZ metric is given by:
\begin{align}
    g_E=
\begin{pmatrix} x-M&iJ/2&0\\ 
    iJ/2&x&0\\ 
    0&0&D^{-1}
\end{pmatrix},
\qquad
\det g_E=\frac14,\quad \sqrt{\det g_E}=\frac12
\end{align}
where $D=4x^2-4Mx+J^2$. Applying the KSW criterion to the above metric then yields for $q=1$ \cite{Basile:2023ycy}: 
\begin{align}
   \mathrm{Re}\left(\sqrt{\det g_E}\,g_E^{-1}\right)
=
\begin{pmatrix}
2x/D&0&0\\
0&2(x-M)/D&0\\
0&0&D/2
\end{pmatrix}.
\end{align}
which is positive definite when: 
\begin{align}
    D>0,\quad x>0, \quad x-M>0.
\end{align}
whereas $q=0,3$ are satisfied and $q=2$ yields the same condition. Thus, in order to satisfy the KSW criterion, the range of the $x$ coordinate must satisfy: 
\begin{align}
    x>\mathrm{max}(0,M)
\end{align}
For the full maximal quotient overspinning geometry with $M>0$, the region $x<M$ fails the criterion, while for the spinning defect geometries with $M<0$, the region containing CTCs $-\beta_-^2\leq x\leq 0$ fails the criterion. It is interesting to see that although CTCs in the Lorentzian geometry containing pathologies do not by themselves exclude the admissibility to the Euclidean path integral, the strict KSW criterion still excludes these continuations.

It is also interesting to note that the de Sitter analogue of the spinning defect geometry, the Kerr-de Sitter spacetime, has been used to compute the partition function of dS$_3$ with an observer \cite{Blommaert:2026ofx}. The authors also retain the region of CTCs in the Lorentzian geometry, placing the observer in the region $r^2<0$. The Lorentzian metric is almost the same as the BTZ metric, except now with a positive cosmological constant and includes a cosmological horizon: 
\begin{align}
    ds^2=-N^2dt^2+\frac{dr^2}{N^2} +r^2\left(d\varphi-\frac{J}{2r^2}dt\right)^2, \qquad N^2=E-r^2+\frac{J^2}{4r^2}
\end{align}
where $E=a^2-b^2$, and $J=2ab$. The observer worldline is placed at $r^2=-b^2$, and the region $-b^2<r^2<0$ contains CTCs. The quasi-Euclidean continuation of the Kerr-de Sitter metric also fails the strict KSW criterion in the CTC region containing the observer. To see it explicitly, it is useful to perform the following coordinate transformation \cite{Blommaert:2026ofx}: 
\begin{align*}
r^2=a^2\sin^2\chi-b^2\cos^2\chi,\quad t=-i\,\frac{a}{a^2+b^2}\,\psi,\quad \varphi=\vartheta-i\,\frac{b}{a^2+b^2}\,\psi.
\end{align*}
and the quasi-Euclidean metric is given by: 
\begin{align}
    ds_E^2=d\chi^2 +a^2\sin^2\chi\,d\vartheta^2 +\cos^2\chi(d\psi+ib\,d\vartheta)^2 
\end{align}
where the radial coordinate retains the CTC region so $0<\chi<\pi/2$. For the real one-form $d\psi$, the KSW criterion gives: 
\begin{align}
    \mathrm{Re}\left(\sqrt g\,g^{\psi\psi}\right) = \frac{a^2\sin^2\chi-b^2\cos^2\chi}{a\sin\chi\cos\chi}>0
\end{align}
The denominator is always positive in the range $0<\chi<\pi/2$, while the numerator is nothing but $r^2$, which violates the KSW criterion precisely in the range $-b^2<r^2<0$, just as for the spinning defect.

So what does this imply for our bulk geometries? If we want to follow the strict KSW criterion, we should end the spacetime at $x=M$ for overspinning geometries and $x=0$ for the spinning defects. This is indeed motivating from the Lorentzian perspective, since both excise the region containing CTCs. However, ending the spacetime will require boundary conditions and matter completions that run into the difficulties described in the previous sections. Moreover, the overspinning geometry needs to be ended at the radial point dependent on its mass, which seems arbitrary from a geometric viewpoint. On the other hand, the KSW criterion seems to be rather strict from a bulk saddle perspective. Other than the Kerr-de Sitter metric above, many useful semiclassical observables have been obtained from complex saddles that fail the KSW criterion \cite{Saad:2018bqo,Chen:2023hra,Bah:2022uyz,Fumagalli:2024msi,Jung:2025fbt}. This has led the authors of \cite{Caminiti:2026efx} to consider a weaker criterion for complex saddles. Nevertheless, we view our bulk interpretations as reasonable proposals whose contribution to the Euclidean path integral remains to be established. 

\section{Discussion}
\label{section4}
In this paper, we revisited the proposal of adding states whose spin scales with the central charge to cure the known negativities in the MWK partition function. We found that in addition to the massless spinning states added in \cite{DiUbaldo:2023hkc}, sub-extremal and extremal spinning states below the black hole threshold, as well as certain massive overspinning states above the black hole threshold can also cure the existing negativities, without introducing new ones. For the sub-extremal and extremal spinning states, there is no upper bound coming from the scalar sector on the degeneracy of states $d$, while for overspinning states, the scalar density can introduce new negativities, and this in turn will place an upper bound on the degeneracy $d$ of these states, which we summarize in Table \ref{summarizedresults}. 

We interpreted the bulk geometries as spinning defects and overspinning BTZ geometries from a minimal matter content perspective. In the case of the spinning defect, the geometry is supported by a delta function source surrounded by a region of CTCs, while the overspinning geometries are smooth saddles without singularities or matter sources, but nonetheless contain causal pathologies. The quasi-Euclidean continuations of these geometries do not satisfy the KSW criterion, and we specified the region which fails it. Since the KSW criterion seems rather strict, we would still like to view our bulk interpretations as reasonable proposals whose contributions to the Euclidean path integral in AdS$_3$ gravity requires further study.

\paragraph{Uniqueness of the spinning states.}
Here we have demonstrated that each of the spinning sectors contains states that are able to cure the existing negativities, without introducing new ones. In particular, any spinning state with $H\in \p{\frac{3}{4}\xi,\xi}$ and $\bar{H}=\frac{3}{4}\xi$ is able to cure the negativities, and there is no upper bound on the degeneracy of states $d$ we may add. The search for overspinning states is guided by the requirement of avoiding new negativities in the scalar density in the regime of $\xi t\lesssim O(1)$, while not imposing overly restrictive conditions on $\xi$. Different spins impose different quantization on the central charge $\xi$, and the choice of which state to add is certainly not fixed by a bottom-up approach. Even if we insist on preserving the spectral gap and restrict to the overspinning states, that choice will not be unique. Thus, it would be interesting to understand if there are additional underlying principles, such as the quantization condition of $\xi$, that might guide the choices of the states one adds in the gravitational path integral, as well as the degeneracy $d$. 

\paragraph{One-loop determinant of spinning geometries.}
The bulk computation of the one-loop determinant on a static conical defect was performed in \cite{Benjamin:2020mfz}, using the technologies developed in \cite{Giombi:2008vd}, where the authors found that the $1/c$ correction for the defect geometries does \textit{not} match the operators they added. This also hints at the fact that perhaps the pure AdS$_3$ quantum gravity does not exist, or perhaps the dual CFT should be viewed as an ensemble average. In any case, it would be interesting to perform similar computations on the spinning defect and overspinning backgrounds, and check whether these match the states we added in the path integral.

\paragraph{Thermodynamics of overspinning geometries and JT gravity?}
Despite the lack of an event horizon, the overspinning geometries carry formally a real chiral temperature, due to the hyperbolic sector of the geometry. It would thus be interesting to understand whether the chiral temperature has any genuine thermodynamic interpretation. Dimensionally reducing the overspinning geometries, as well as their modular images, to JT gravity might be able to help clarify such questions.

\acknowledgments

I would like to thank David Berenstein, Steve Carlip, and Jacqueline Caminiti for helpful discussions. I would also like to acknowledge Claude and ChatGPT for assistance with numerical computations, TikZ figure preparation, and language/format editing.

\appendix

\section{Review of the MWK partition function}
\label{appenA}
In this appendix, we review the basic assumptions and computations of the Euclidean gravitational path integral in three-dimensional anti-de Sitter spacetime, and record several useful formulae. The action of three-dimensional Einstein gravity is given by: 
\begin{align}
\label{actionads3}
I_E=-\frac{1}{16\pi G}\int_{\mathcal M} d^3x\,\sqrt{g}\,\left(R+\frac{2}{l^2}\right)
-\frac{1}{8\pi G}\int_{\partial\mathcal M} d^2x\,\sqrt{\gamma}\,\left(K-\frac{1}{l}\right)
\end{align}
where $G$ is the three-dimensional Newton's constant, $R$ is the Ricci scalar, $K$ is the extrinsic curvature, and $\g$ is the induced metric on the boundary. We will set the AdS$_3$ radius $l=1$. The main goal is then to perform the gravitational path integral by integrating over all metric degrees of freedom: 
\begin{align}
    Z=\int_{\partial\mathcal{M}=T^2}\mathcal{D}g\,e^{-I_E\left[g\right]}
\end{align}
There is no known way to perform the above integral exactly, even in the simple setting of three-dimensional gravity. Maloney and Witten argued that one can in fact perform the above integral by a saddle point approximation, which sums over all known smooth on-shell Euclidean saddles with torus boundary condition \cite{Maloney:2007ud}: 
\begin{align}
    Z\overset{!}{=}\sum_{g_0}\,e^{-I_E\left[g_0\right]}
\end{align}
Related fixed-topology formulations based on Virasoro TQFT have since been proposed in \cite{Collier:2023cyw,Collier:2024fid}. The simplest saddle that satisfies the above criterion is the Euclidean AdS$_3$ geometry: 
\begin{align}
\label{eads3metric}
    \dd s^2=(1+r^2)\dd t_E^2+\frac{\dd r^2}{1+r^2}+r^2 \dd \phi^2
\end{align}
with torus boundary parametrized by $(t_E,\phi)\sim (t_E,\phi+2\pi)\sim (t_E+\beta,\phi+\theta)$. The classical action of Euclidean AdS$_3$ can be easily evaluated using (\ref{actionads3}), which is given by:
\begin{align}
    I_{\text{EAdS}_3}=-\frac{\beta}{8G}
\end{align}
It is useful to define the modular parameters
\begin{align}
    \tau=\frac{\theta+i\beta}{2\pi},\quad \bar \tau=\frac{\theta-i\beta}{2\pi}
\end{align}
so that the semi-classical partition function of Euclidean AdS$_3$ can be written as: 
\begin{align}
\label{eads3}
    Z_{\text{EAdS}_3}(\tau,\bar \tau)\approx q^{-\frac{1}{16G}}\bar q^{-\frac{1}{16G}}, \qquad
q=e^{2\pi i\tau},
\quad
\bar q=e^{-2\pi i\bar\tau}.
\end{align}
This in fact has a simple interpretation in terms of the dual CFT$_2$ with central charge $c=\frac{3}{2G}$ \cite{Brown:1986nw}, where the above formula is simply the classical contribution to the partition function of the vacuum state: 
\begin{align}
    Z(\beta,\theta)=\Tr\!\left[e^{-\beta \mathcal H+i\theta \mathcal J}\right]=\Tr\!\left[ q^{\,L_0-\frac{c}{24}} \bar q^{\,\bar L_0-\frac{c}{24}}
\right]
\end{align}
with eigenvalues $L_0=\bar L_0=0$. Note that here we have defined the Hamiltonian and the angular momentum as: 
\begin{align}
    \mathcal  H=L_0+\bar L_0-\frac{c}{12},\qquad J=L_0-\bar L_0
\end{align}
Since there are no local propagating degrees of freedom, one might have hoped that (\ref{eads3}) is an exact answer. This, however, is not true. There are in fact boundary gravitons, or simply the descendant states of the vacuum state given by: 
\begin{align}
    \prod_{n=2}^{\infty}L_{-n}^{N_n}\,
    \prod_{m=2}^{\infty}\bar L_{-m}^{\bar N_m}\,|0\rangle,
    \qquad
    N_n,\bar N_m\in \mathbb Z_{\ge 0}.
\end{align}
Note that the products here start from $m,n=2$ since $L_{-1},\bar L_{-1}$ annihilate the CFT vacuum state. The full contribution to the partition function is then: 
\begin{align}
    Z_{\text{EAdS}_3}(\tau,\bar \tau)=q^{-\frac{c}{24}}\bar q^{\, -\frac{c}{24}}\frac{1}{\prod_{n=2}^\infty\abs{1-q^n}^2}=\frac{1}{\abs{\eta\p{\tau}}^2}q^{-\xi}\bar q^{\, -\xi}\abs{1-q}^2
\end{align}
where we have introduced the Dedekind $\eta$ function:
\begin{align}
    \eta(\tau)=q^{\frac{1}{24}}\prod_{n=1}^{\infty}\p{1-q^n}.
\end{align}
and defined $\xi=\frac{c-1}{24}$. The above contribution is one-loop exact, and one expects that higher loop contributions will only renormalize the central charge\cite{Maloney:2007ud}. It is often useful to isolate the modular invariant piece $\sqrt{\text{Im}\t}\abs{\eta(\tau)^2}$ and write the partition function as: 
\begin{align}
    Z_{\text{EAdS}_3}(\tau,\bar \tau)=\frac{1}{\sqrt{\text{Im}\t}\abs{\eta(\tau)^2}}\left[\sqrt{\text{Im}\t}\,q^{-\xi}\bar q^{\, -\xi}\abs{1-q}^2\right]
\end{align}
Evidently, Euclidean AdS$_3$ is not the only saddle that contributes to the path integral; there are black hole solutions discovered by Ba\~nados, Teitelboim, and Zanelli \cite{Banados:1992wn}, which should also contribute, since their Euclidean continuations are smooth on-shell saddles with a torus boundary. In fact, the Euclidean BTZ geometry can be obtained by a simple $s$-modular transformation $\tau \to -\frac{1}{\tau}$ of (\ref{eads3metric}), and the partition function is simply given by:
\begin{align}
    Z_{\text{BTZ}}\p{\tau',\bar \tau'}= Z_{\text{EAdS}_3}(-\frac{1}{\tau},-\frac{1}{\bar \tau})
\end{align}
It is then natural to expect that by acting with the full modular group $\g\in SL(2,\mathbb{Z})$: 
\begin{align}
    \g \cdot\tau=\frac{a\tau+b}{s\tau+d}\,\qquad
a,b,s,d\in\mathbb Z,\quad ad-bs=1
\end{align}
one can then generate all classified Euclidean saddles that contribute to the gravitational path integral. This is almost true, except that the $T-$transformation $\tau \to \tau+1$ will not generate new geometries, and we need to exclude these transformations so that we are not over-counting the contributions. This led the authors of \cite{Maloney:2007ud} to consider the following three-dimensional gravitational path integral: 
\begin{align}
    Z(\tau,\bar \tau)=\sum_{\g\in \G_\infty\setminus PSL(2,\mathbb{Z})}  Z_{\text{EAdS}_3}(\g\cdot\tau, \g\cdot\bar \tau)
\end{align}
where $ \G_\infty$ is the subgroup of $PSL(2,\mathbb{Z})$ that acts with $\tau \to \tau+n,\hspace{1mm}n\in \mathbb{Z}$. This means that when performing the sum, we only need to consider cases when $s,d_\g$ are coprime, since $ad_\g-bs=1$. The geometries generated by $ \g\in  \G_\infty\setminus PSL(2,\mathbb{Z}) $ labeled by $s,d_\g$ are known as the $SL(2,\mathbb{Z})$ black holes, and in general they do not have a simple Lorentzian interpretation as the usual BTZ black hole but nonetheless serve as admissible Euclidean saddle points. 

\subsection{Density of states}
To compute the density of states, it is useful to write the partition function in terms of the Virasoro character of a primary state with weight $H$: 
\begin{align}
    \chi_H(\tau)=\frac{q^{H-\xi}}{\eta(\tau)}
\end{align}
where the vacuum character can be written as: 

\begin{align}
\label{eadspar}
    Z_{\text{EAdS}_3}(\tau,\bar \tau)=&\abs{\chi_{\text{vac}}(\tau,\bar \tau)}^2 \notag \\
    =&\frac{1}{\abs{\eta\p{\tau}}^2}\abs{q^{-\xi}(1-q)}^2=\p{\chi_0(\tau)-\chi_1(\tau)}\p{\bar{\chi}_0(\bar \tau)-\bar{\chi}_1(\bar\tau)}
\end{align}

We can use the modular crossing kernel $\mathbb{K}_{h,H}^{\g}$ to express the modular transform of a Virasoro character $\chi_H(\g \t)$ as \cite{Benjamin:2020mfz}: 
\begin{align}
    \chi_H(\g \t)=\int_\xi^\infty \dd h \hspace{1mm}\mathbb{K}_{h,H}^{\g} \chi_h(\t)
\end{align}
where $\g  \in PSL(2,\mathbb{Z})$. The explicit form of the modular crossing kernel for $s>0$ is given by: 
\begin{align}
    \mathbb{K}_{h,H}^{\g}=\epsilon(\g)\sqrt{\frac{2}{s}}e^{\frac{2\pi i}{s}\p{a\cdot \p{H-\xi}+d\cdot \p{h-\xi}}}\frac{\cos\p{\frac{4\pi }{s}\sqrt{\p{H-\xi}\p{h-\xi}}}}{\sqrt{\p{h-\xi}}}
\end{align}
where $\epsilon(\g)$ is a phase factor that only depends on the weights. The density of states of a seed state with conformal weights $(H, \bar H)$ is then simply the $PSL(2,\mathbb{Z})$ sum of the modular crossing kernels: 
\begin{align}
  Z(\tau,\bar \tau)=\!\sum_{\g\in \G_\infty\setminus PSL(2,\mathbb{Z})}\abs{\chi_H(\g\tau) }^2\implies \hspace{2mm}\rho(H,\bar H)=\!\sum_{\g\in \G_\infty\setminus PSL(2,\mathbb{Z})} \mathbb{K}_{h,H}^{\g} \overline{\mathbb{K}}_{\bar h,\bar H}^{\g}
\end{align}
Similarly, from (\ref{eadspar}) we can write the density of states of the MWK partition function as: 
\begin{align}
    \rho^{\text{MWK}}\p{h,\bar h}=\!\sum_{\g\in \G_\infty\setminus PSL(2,\mathbb{Z})}\left[
\mathbb{K}^{(\gamma)}_{h0}\,\overline{\mathbb{K}}^{(\gamma)}_{\bar h 0}
-
\mathbb{K}^{(\gamma)}_{h0}\,\overline{\mathbb{K}}^{(\gamma)}_{\bar h 1}
-
\mathbb{K}^{(\gamma)}_{h1}\,\overline{\mathbb{K}}^{(\gamma)}_{\bar h 0}
+
\mathbb{K}^{(\gamma)}_{h1}\,\overline{\mathbb{K}}^{(\gamma)}_{\bar h 1}
\right]
\end{align}
Using the explicit form of the modular crossing kernel, one can manipulate the above sum and decompose it into spin sectors. This is done explicitly in appendix A of \cite{Benjamin:2020mfz}. Restricting the sum to $s>0$, it is useful to decompose the sum over $d$ by rewriting $d=d_0+sn$, where $d_0\in \p{\mathbb{Z}/s\mathbb{Z}}^*$, and sum over $d_0$ and $n\in \mathbb Z$ separately. Note also that $ad-1=bs$ gives $ a=d^{-1} \mod{s}$. Thus, we have: 
\begin{align}
    \!\sum_{\substack{\g\in \G_\infty\backslash PSL(2,\mathbb{Z})\\s(\g)>0}}=\sum_{s>0}\sum_{d_0\in \p{\mathbb{Z}/s\mathbb{Z}}^*} \sum_{n\in \mathbb Z}.
\end{align}
The density of states can be worked out term by term. For example: 
\begin{align}
    &\sum_{s>0}\sum_{d_0\in \p{\mathbb{Z}/s\mathbb{Z}}^*} \sum_{n\in \mathbb Z}\mathbb{K}^{(\gamma)}_{h0}\,\overline{\mathbb{K}}^{(\gamma)}_{\bar h 1} \notag \\
    &=\frac{2}{\sqrt{t \bar t}}\sum_{s>0}\sum_{n\in \mathbb Z} \frac{S(j,-1;s)}{s} e^{2\pi i nj}\cos\p{\frac{4\pi}{s}\sqrt{-\xi\p{h-\xi}}}\cos\p{\frac{4\pi}{s}\sqrt{(1-\xi)\p{\bar h-\xi}}} \notag \\
    &=\frac{2}{\sqrt{t \bar t}}\sum_{s>0}\sum_{\ell \in \mathbb Z} \frac{S(j,-1;s)}{s} \delta\p{j-\ell}\cos\p{\frac{4\pi}{s}\sqrt{-\xi\bar t}}\cos\p{\frac{4\pi}{s}\sqrt{(1-\xi) t}}
\end{align}
where we assumed $j\geq0$ ($t=\text{min}(h,\bar h)-\xi$)\footnote{Note that the MWK density of states (\ref{MWK partition function}) is invariant under $j\to -j$, since $S(j,1;s)=S(-j,-1;s)$. }, and $S(j,J,s)$ is the Kloosterman sum: 
\begin{align}
S(j,J;s)=\sum_{\substack{0\le d<s,\hspace{1mm} \gcd(s,d)=1}}
\exp\!\left(
2\pi i \,\frac{dj + (d^{-1})_s J}{s}
\right),
\end{align}
The rest of the terms can be similarly computed, and decomposing into spin-sectors gives: 
\begin{align}
\label{MWK partition function}
   \rho_j^{\text{MWK}}=\frac{2}{\sqrt{t \bar{t}}}\sum_{s=1}^{\infty}\frac{1}{s}&\left[S(j,0;s)\cosh\p{\frac{4\pi}{s}\sqrt{\xi \bar t}}\cosh\p{\frac{4\pi}{s}\sqrt{\xi t}} \right. \notag\\ 
   &-S(j,-1;s)\cosh\p{\frac{4\pi}{s}\sqrt{\xi \bar t}}\cosh\p{\frac{4\pi}{s}\sqrt{\p{\xi-1} t}} \notag \\
   &-S(j,1;s)\cosh\p{\frac{4\pi}{s}\sqrt{\p{\xi-1} \bar t}}\cosh\p{\frac{4\pi}{s}\sqrt{\xi t}} \notag \\
   &\left. +S(j,0;s)\cosh\p{\frac{4\pi}{s}\sqrt{\p{\xi-1} \bar t}}\cosh\p{\frac{4\pi}{s}\sqrt{\p{\xi-1} t}} \right]
\end{align}
The above sum is divergent, and requires regularization. One might worry that different regularization schemes can lead to different results, but it has been shown that they all lead to the same negative density of states at the threshold \cite{Maloney:2007ud,Keller:2014xba,Stanford:2025llj}. One way to regularize the sum is to introduce a regularized crossing kernel with a generic weight $w$, where $w=\frac{1}{2}$ recovers the weight of interest. By isolating the divergent piece first, performing the sum over $s$, and then taking the $w\to\frac{1}{2}$ limit, the regularized sum at the threshold is given by \cite{Maloney:2007ud,Keller:2014xba,Benjamin:2020mfz}: 
\begin{align}
    \tilde{\rho}^{\text{MWK}}_0(t)=-6\delta(t)+\frac{2}{t}\sum_{s=1}^{\infty} \biggl\{\frac{\phi(s)}{s}\left[\sinh^2\p{\frac{4\pi}{s}\sqrt{\xi t}}+\sinh^2\p{\frac{4\pi}{s}\sqrt{(\xi-1)t}}\right] \notag \\
    -2\frac{\mu(s)}{s}\left[\cosh\p{\frac{4\pi}{s}\sqrt{\xi t}}\cosh\p{\frac{4\pi}{s}\sqrt{(\xi-1)t}}-1\right]\biggl\}
\end{align}
where the delta function arises from integrating $t$ in a finite range from 0 to a positive number and then taking the $w\to \frac{1}{2}$ limit. With the same method, one can work out the scalar density for a seed state with $T<0$ and $J>0$:
\begin{align}
    \tilde{\rho}_{j=0}^{(T,J)}(t)=2\sigma_0(J)\delta(t)+\sum_{s=1}^{\infty}\frac{2S(0,J;s)}{st}\left[\cosh\p{\frac{4\pi}{s}\sqrt{-Tt}}\cosh\p{\frac{4\pi}{s}\sqrt{-\bar{T}t}}-1\right]
\end{align}

\section{Pole structure of overspinning correlator}
\label{overspinpoles}
Here we describe the $i\epsilon$-prescription for the Lorentzian image correlator for the overspinning geometry. In particular, we find that the complex frequency branch is not the actual pole of the image correlator, which can be written as: 
\begin{equation}
\label{eq:os-wightman-images}
   G^+_\epsilon(t,\phi) = \sum_{m\in\mathbb Z} \left[ \frac{a}{\sinh a(\phi-t+2\pi m+i\epsilon)} \right]^\Delta
   \left[ \frac{b}{\sin b(\phi+t+2\pi m-i\epsilon)}\right]^\Delta .
\end{equation}
The regulator follows from the $t\to t-i\epsilon$ prescription. For non-integer conformal dimension $\D$, the expansion also needs a branch prescription, which we write for the hyperbolic sector:
\begin{equation}
\left[\frac{a}{\sinh az}\right]^\Delta = \exp\!\left\{ \Delta\left[\log a-\operatorname{Log}(\sinh az)\right] \right\}
\end{equation}
where $\mathrm{Log}$ is defined as picking the principal logarithm. Similarly, the sine factor can be expanded using the binomial theorem, where we have:
\begin{align}
\label{eq:os-sine-branch}
\left[\frac{b}{\sin bz}\right]^\Delta &=
(2b)^\Delta e^{i\pi\Delta/2} e^{-ib\Delta z} \left(1-e^{-2ibz}\right)^{-\Delta}
\nonumber\\
&= (2b)^\Delta e^{i\pi\Delta/2} \sum_{k=0}^{\infty}c_k e^{-iq_k z}
\end{align}
where $c_k=\Gamma(k+\Delta)/\p{\Gamma(\Delta)\Gamma(k+1)}$, and $q_k=b(2k+\D)$. Since $\ell$ is quantized, we can write the Fourier coefficient of the propagator as: 
\begin{align}
\label{eq:os-unfolded}
G^+_{\ell,\epsilon}(t)
&=
\frac{1}{2\pi}\int_0^{2\pi}
d\phi\,e^{-i\ell\phi}G^+_\epsilon(t,\phi)
\nonumber\\
&=
\frac{1}{2\pi}\int_{\mathbb R}dy\,e^{-i\ell y}
\left[\frac{a}{\sinh a(y-t+i\epsilon)}\right]^\Delta
\left[\frac{b}{\sin b(y+t-i\epsilon)}\right]^\Delta .
\end{align}
where we have defined $y=\phi+2\pi m$ and used the fact that $e^{2\pi i m\ell}=1$. We can get rid of the time dependence by defining $s=y-t$ and since both $y$ and $t$ are real, the integration range is unchanged: 
\begin{equation}
G^+_{\ell,\epsilon}(t)
=
\frac{(2b)^\Delta e^{i\pi\Delta/2}}{2\pi}
\sum_{k=0}^{\infty}
c_k e^{-q_k\epsilon}
I_\epsilon(p_{k\ell})e^{-i\Omega_{k\ell}t},
\end{equation}
where $p_{k\ell}=\ell+q_k$ and $\Omega_{k\ell}=\ell+2q_k$. The integral is: 
\begin{align}
    I_\epsilon(p)=
\int_{\mathbb R}ds\,e^{-ips}
\left[\frac{a}{\sinh a(s+i\epsilon)}\right]^\Delta
\end{align}
To perform the integral, it is useful to move the contour $z=s+i\epsilon$ to $\mathrm{Im}z=\pi/2a$. Since no poles are crossed, the integral can be evaluated as: 
\begin{align}
I_\epsilon(p) &=
e^{-p\epsilon-i\pi\Delta/2} e^{\pi p/(2a)} a^\Delta \int_{\mathbb R}ds\, \frac{e^{-ips}}{(\cosh as)^\Delta}
\notag\\
&= e^{-p\epsilon-i\pi\Delta/2} \frac{(2a)^{\Delta-1}}{\Gamma(\Delta)} e^{\pi p/(2a)} \left| \Gamma\!\left(\frac{\Delta}{2} +\frac{ip}{2a}\right) \right|^2 
\label{eq:os-hyperbolic-transform}
\end{align}
where in the second line we have used the Euler beta integral. Plugging this back to the Fourier mode, we have: 
\begin{equation}
\label{eq:os-wightman-modes}
G^+_{\ell,\epsilon}(t) = \sum_{k=0}^{\infty} A_{k\ell} e^{-i\Omega_{k\ell}t-\epsilon\Omega_{k\ell}},
\qquad
\Omega_{k\ell}=\ell+2b(2k+\Delta),
\end{equation}
with a rather complicated expression for $A_{k\ell}$:
\begin{equation}
\label{eq:os-mode-weights}
A_{k\ell} = \frac{(2b)^\Delta(2a)^{\Delta-1}}{2\pi} \frac{\Gamma(k+\Delta)}
{\Gamma(\Delta)^2\Gamma(k+1)} e^{\pi p_{k\ell}/(2a)}
\left| \Gamma\!\left( \frac{\Delta}{2}+\frac{ip_{k\ell}}{2a} \right) \right|^2 
\end{equation}
Now that we have the mode expansion, we can assemble the Feynman propagator:
\begin{equation}
G_F(t,\phi)
= \theta(t)G^+(t,\phi) +\theta(-t)G^+(-t,-\phi),
\end{equation}
where $G^+=\lim_{\epsilon\to0^+}G^+_\epsilon$. The Fourier coefficient of $e^{-i\ell \phi}$ is thus given by:
\begin{equation}
G_{F,\ell}(t) = \theta(t)G^+_\ell(t) +\theta(-t)G^+_{-\ell}(-t).
\end{equation}
To extrapolate the poles, we Fourier transform in time with $e^{-i\omega t}$ where we regulate the integral with $e^{-\epsilon t}$. This yields: 
\begin{equation}
\label{eq:os-feynman-spectrum}
G_{F,\ell}(\omega)
= i\sum_{k=0}^{\infty}
\left[ \frac{A_{k\ell}}{\omega-\Omega_{k\ell}+i\epsilon} - \frac{A_{k,-\ell}}{\omega+\Omega_{k,-\ell}-i\epsilon}\right],
\end{equation}
Indeed, we find the two families of poles related by $\ell \to -\ell$, and are given by: 
\begin{equation}
\omega=\ell+2b(2k+\Delta),
\qquad
\omega=\ell-2b(2k+\Delta) 
\end{equation}
In particular, they do not contain the $\omega_R=-\ell-2ia(2k+\D)$ complex frequency modes for the specified image prescription, which should have already been clear in the mode expansion (\ref{eq:os-wightman-modes}), where the frequency only contains the real normal-modes. Although the mode expansion obtained here is not a convergent series for $\Delta>1$, one can introduce a regulator that is polynomial in $\omega$, and does not affect the poles.

\section{Extremal defect correlator}
\label{extremalcorrelator}
The metric can be written as: 
\begin{align}
    \dd s^2=-\p{r^2+2\beta_+^2}\dd t^2+ \frac{r^2 \dd r^2}{\p{r^2+\beta_+^2}^2}+r^2 \dd \phi^2 -2\beta_+^2\dd \phi \hspace{1mm} \dd t
\end{align}
We will study the massive scalar field perturbations on this background. It is useful to implement the following coordinate transformation: 
\begin{align}
    z=\frac{\beta_+^2}{r^2+\beta^2_+}
\end{align}
and the metric is given by: 
\begin{align}
\label{transformedmetric}
    \dd s^2=-\frac{1+z}{z}\beta_+^2 \dd t^2+\frac{\dd z^2}{4z^2}-2\beta_+^2 \dd t \dd \phi +\frac{1-z}{z} \beta_+^2 \dd \phi^2
\end{align}
The wave equation is given by: 
\begin{align}
    \p{\nabla^2-m^2}\Phi(\textbf{x})=0
\end{align}
where we use the ansatz $\Phi\p{\textbf{x}}=e^{-i\omega t}e^{i\ell \phi}R(z)$. In these coordinates (\ref{transformedmetric}), the wave equation is the Whittaker equation: 
\begin{align}
    \frac{d^2 R(z)}{dz^2}-R(z)\p{\frac{\p{z-1}\omega^2}{4z \beta_+^2}+\frac{\ell \omega}{2\beta_+^2}+\frac{\p{1+z}\ell^2}{4z \beta_+^2}+\frac{\Delta\p{\Delta-2}}{4z^2}}=0
\end{align}
The solution is given by the Whittaker function (confluent hypergeometric function), and has two singularities at $z=0$ and $z=\infty$. This can be thought of as a limiting case of the hypergeometric function, where two of the three regular singularities merge into an irregular singularity. In this case, we have two linearly independent solutions: 
\begin{align}
    R_1(z)=M_{\frac{\omega -\ell}{4 \beta _+},\frac{\Delta -1}{2}}\left(\frac{z (\ell+\omega )}{\beta _+}\right), \qquad
    R_2(z)=W_{\frac{\omega -\ell}{4 \beta _+},\frac{\Delta -1}{2}}\left(\frac{z (\ell+\omega )}{\beta _+}\right)
\end{align}
These can be recast into the confluent hypergeometric function by the following identities: 
\begin{align}
    M_{k,m}(z)&=e^{-\frac{z}{2}}z^{\frac{1}{2}+m}F\p{\frac{1}{2}+m-k,1+2m;z} \notag \\
    W_{k,m}(z)&=e^{-\frac{z}{2}}z^{\frac{1}{2}+m}U\p{\frac{1}{2}+m-k,1+2m;z}    
\end{align}
where 
\begin{align}
    U(a,b;z)=\frac{\G\p{1-b}}{\G\p{1-b+a}}F\p{a,b;z}+\frac{\G\p{b-1}}{\G\p{a}}z^{1-b}F\p{1+a-b,2-b;z}
\end{align}
We impose a boundary condition at the location of the defect $z=\infty$ $(r^2=-\beta_+^2)$, where we require the solution to behave regularly. The two functions behave differently at the two singularities: 
\begin{align}
     R_1(z) =
    \begin{cases}
      z\to \infty \hspace{1mm} (r^2=-\beta_+^2)& \text{Diverges if $\ell+\omega>0$}\\
      z\to0 \hspace{1mm} (r^2\to \infty) & \text{Regular }\\
    \end{cases}   
\end{align}
Moreover, we have: 
\begin{align}
    R_2(z) =
    \begin{cases}
      z\to \infty \hspace{1mm} (r^2=-\beta_+^2)& \text{Regular if $\ell+\omega>0$}\\
      z\to0 \hspace{1mm} (r^2\to \infty) & \text{Diverges }\\
    \end{cases}  
\end{align}
Imposing regularity at $z=\infty$ selects $W_{k,m}$, where the function vanishes under the condition $\ell+\omega>0$. Next, we study this solution's behavior at the boundary $z=0(r=\infty)$: 
\begin{align}
\label{ayms}
R_2(z)\sim &\left(\frac{z (\ell+\omega )}{\beta _+}\right){}^{-\Delta /2}\p{\frac{(\ell+\omega ) \Gamma (\Delta -1) z}{\Gamma \left(\frac{\ell-\omega +2 \Delta  \beta _+}{4 \beta _+}\right) \beta _+}+O\left(z^2\right)}+ \notag \\
   &+\left(\frac{ z (\ell+\omega )}{\beta_+}\right){}{}^{\Delta /2}\p{\frac{\Gamma (1-\Delta )}{\Gamma \left(\frac{\ell-\omega -2 \Delta  \beta _++4 \beta _+}{4 \beta _+}\right)}+O\left(z\right)}
\end{align}
We see that the first term will diverge at $z=0$, which yields the quantization condition: 
\begin{align}
    \frac{\ell-\omega +2 \Delta  \beta _+}{4 \beta _+}=-k, \quad k\in \mathbb{Z}_{\geq0}
\end{align}
This implies: 
\begin{align}
    \omega_{k,\ell}=\ell+2\beta_+\p{2k+\Delta}, \quad k\in \mathbb{Z}_{\geq0}
\end{align}
Note that condition $Q=\omega+\ell>0$ is satisfied, meaning that $p_{k\ell}=\ell+\beta_+(2k+\Delta)>0$. The other branch is given by $Q<0$ with $p_{k\ell}<0$. Plugging this back into our solution, we have: 
\begin{align}
    &R_2(z)=W_{\frac{\omega -\ell}{4 \beta _+},\frac{\Delta -1}{2}}\left(\frac{z (\ell+\omega )}{\beta _+}\right)=\p{\frac{\omega+\ell}{\beta_+}}^{\Delta/2}e^{-z\frac{\ell+\omega}{2\beta_+}}z^{\Delta/2}U\left(-k,\Delta;\frac{z(\omega+\ell)}{\beta_+}\right) \notag \\*
    &U\left(-k,\Delta;\frac{z(\omega+\ell)}{\beta_+}\right) =\frac{\G\p{1-\Delta}}{\Gamma\p{1-\Delta-k}}F\p{-k,\Delta;\frac{z(\omega+\ell)}{\beta_+}}
\end{align}
where we have dropped the second term since we have in the denominator $\G\p{-k}$. The confluent hypergeometric function with the first entry being a negative integer reduces to generalized Laguerre polynomials, which obey the respective orthogonality condition. This can be used to compute the usual inner product:
\begin{align}
\label{normm}
\langle \Phi_I,\, \Phi_J \rangle &= -i \int_{\Sigma} \dd^2 x \sqrt{g_{\Sigma}(x)}\, n^{\mu} \Phi_I(x) \overset{\leftrightarrow}{\partial}_{\!\mu} \Phi_J^*(x) =\delta_{IJ}
\end{align}
Here $(k,\ell)$ labels the modes, and $n^\mu$ denotes the unit norm of the constant time-slice. We should again emphasize that the above inner product is not the standard one defined on a Cauchy surface, but an analytic continuation that retains the CTC region. The norm vector is: 
\begin{align}
    n^{a}=\p{-\frac{g_{\phi \phi}}{g_{t \phi}}b\hspace{1mm},\hspace{1mm} 0\hspace{1mm},\hspace{1mm} b}, \hspace{6mm} b=\pm \p{-g_{tt}\p{\frac{g_{\phi \phi}}{g_{t\phi}}}^2+g_{\phi \phi}}^{-1/2}
\end{align} 
We take $b>0$ so that the final inner product is positive. Thus, we have:
\begin{align}
\label{ints}
    \langle \Phi_I,\, \Phi_J \rangle=2\int_0^{\infty}\dd z\sqrt{g_{\Sigma}} \hspace{1mm}n^t \omega \abs{R_2(z)}^2-2\int_0^\infty\dd z \sqrt{g_{\Sigma}} \hspace{1mm} n^\phi \ell \abs{R_2(z)}^2
\end{align}
where we have: 
\begin{align*}
    \sqrt{g_{\Sigma}} \hspace{1mm}n^t=\frac{1-z}{2z} ,\quad
    \sqrt{g_{\Sigma}} \hspace{1mm}n^\phi=\frac{1}{2}
\end{align*}
Recall that: 
\begin{align}
    R_2(z)=\p{\frac{\omega+\ell}{\beta_+}}^{\Delta/2}e^{-z \frac{\ell+\omega}{2\beta_+}}z^{\Delta/2} \p{-1}^k\G\p{k+1}L_{k}^{\Delta-1}\p{z\frac{\omega+\ell}{\beta_+}}
\end{align}
where $L_{k}^{\Delta-1}$ is the generalized Laguerre polynomials. The first integral in (\ref{ints}) is then given by: 
\begin{align*}
   2 \int_0^{\infty}\dd z\sqrt{g_{\Sigma}} \hspace{1mm}n^t \omega \abs{R_2(z)}^2=\underbrace{\omega\int_0^\infty\dd z \frac{1}{z}\abs{R_2(z)}^2}_{(1)}-\underbrace{\omega\int_0^\infty\dd z \abs{R_2(z)}^2}_{(2)} 
\end{align*}
where 
\begin{align*}
    (1)=&\omega\cdot\p{\frac{\beta_+}{\omega+\ell}}\p{\frac{\beta_+}{\omega+\ell}}^{-1}\G\p{1+k}^2\int_0^\infty \dd x \hspace{1mm} e^{-x} x^{\Delta-1}\abs{L_k^{\Delta-1}(x)}^2 \\
    =&\omega \cdot \G\p{1+k}^2\frac{\G\p{k+\Delta}}{\G\p{1+k}}\\
    =&\omega \cdot  \G(1+k)\G\p{k+\Delta}
\end{align*}
where we have made a change of variables $z=\frac{\beta_+}{\omega+\ell}x$. Similarly $(2)$ can be evaluated as: 
\begin{align*}
    -(2)=&-\omega \cdot \p{\frac{\beta_+}{\omega+\ell}} \G(k+1)^2 \int_0^{\infty}\dd x \hspace{1mm}e^{-x}x^\Delta \abs{L_k^{\Delta-1}(x)}^2 \\*
    =&-\omega \cdot \p{\frac{\beta_+}{\omega+\ell}}\G\p{1+k}\G\p{k+\Delta}\p{2k+\Delta}
\end{align*}
The second integral in (\ref{ints}) is the same as (2), which gives: 
\begin{align*}
    -2\int_0^\infty\dd z \sqrt{g_{\Sigma}} \hspace{1mm} n^\phi \ell \abs{R_2(z)}^2=&-\ell\int_0^\infty \dd z \abs{R_2(z)}^2 \\
    =&-\ell\cdot \p{\frac{\beta_+}{\omega+\ell}}\G(k+1)^2 \int_0^{\infty}\dd x \hspace{1mm}e^{-x}x^\Delta \abs{L_k^{\Delta-1}(x)}^2  \\
    =&-\ell \cdot \p{\frac{\beta_+}{\omega+\ell}}\G\p{1+k}\G\p{k+\Delta}\p{2k+\Delta}
\end{align*}
Thus, combining everything, we have: 
\begin{align*}
     \langle \Phi_I,\, \Phi_I \rangle=&\p{\omega-\beta_+\p{2k+\Delta}}\G\p{1+k}\G\p{k+\Delta} \\
     =&\mathcal{N}_{k\ell}^2
\end{align*}
The above computation applies to the positive-frequency branch $p_{k\ell}>0$. As discussed in the main text, the negative branch can be obtained by $-\ell,-\omega$. The extrapolate dictionary then gives:
\begin{align}
G^+(t,\phi)
&= \lim_{r_1,r_2\to\infty}
r_1^\Delta r_2^\Delta G_{\rm bulk}^+=\beta_+^{2\Delta}\lim_{z_1,z_2\to 0} \! \Big( z_1^{-\Delta/2} z_2^{-\Delta/2} G^+(z_1,z_2) \Big) \notag\\
&= \sum_{k \in \mathbb Z_{\geq0}}\sum_{\substack{\ell\in\mathbb Z\\p_{k\ell}>0}}
e^{-i\omega_{k\ell} t} e^{i\ell\phi}
\Big( \mathcal{N}_{k\ell}^{-1} \psi_{k\ell}(\Delta) \Big)^2,
\end{align}
where from (\ref{ayms}) we have: 
\begin{align}
    \psi_{k\ell}(\Delta)=\p{\frac{\ell+\omega_{k\ell}}{\beta_+}}^{\Delta/2}\frac{\beta_+^{\Delta}\Gamma (1-\Delta ) }{\Gamma (1-k-\Delta )}.
\end{align}
The extra factor of $\beta_+^\D$ comes from the extrapolate dictionary, where $r\sim \beta_+ z^{-1/2}$ near $z=0$. Combining everything, we have: 
\begin{align}
    G^+(t,\phi)&=\sum_{k \in \mathbb Z_{\geq0}}\sum_{\substack{\ell\in\mathbb Z\\p_{k\ell}>0}}
    e^{-i\omega_{k\ell} t} e^{i\ell\phi}
    \Big( \mathcal{N}_{k\ell}^{-1} \psi_{k\ell}(\Delta) \Big)^2, \notag \\
    \Big( \mathcal{N}_{k\ell}^{-1} \psi_{k\ell}(\Delta) \Big)^2 &= \frac{\p{2\beta_+}^\Delta p_{k\ell}^{\Delta -1}\Gamma (1-\Delta )^2}{\Gamma(1+k) \Gamma (1-k-\Delta )^2 \Gamma (k+\Delta )}
\end{align}
This can be simplified as:
\begin{align}
    \frac{ \Gamma (1-\Delta )^2 }{\Gamma(1+k) \Gamma (1-k-\Delta )^2 \Gamma (k+\Delta )}=\frac{\Gamma (k+\Delta )}{\Gamma (\Delta )^2 \Gamma (k+1)}
\end{align}
matching exactly with (\ref{extremalpropagator}).  The $Q<0$ modes then give $G^-(t,\phi)=G^+(-t,-\phi)$.

\subsection{Equivalence between the image expression and the mode sum}
\label{appenb1}
Here we show that the image expression is equivalent to the bulk mode sum. We retain the $i\epsilon$-prescription throughout and take the $\epsilon\to 0^+$ limit in the end. Define $x_m=\phi+2\pi m$, and (\ref{extremalpropa}) can be written as: 
\begin{align}
    G^+_{\mathrm{img},\epsilon}(t,\phi)
    =\beta_+^\Delta\sum_{m\in\mathbb Z}
        (x_m-t+i\epsilon)^{-\Delta}
        \left[\sin\p{\beta_+(x_m+t-i\epsilon)}\right]^{-\Delta}
\end{align}
The above regulated sum converges absolutely for $\Delta>1$. Using the binomial theorem we can expand the sine factor as: 
\begin{align}
    (\sin u)^{-\Delta}
    =(2i)^\Delta\sum_{k=0}^\infty
        \frac{\Gamma(\Delta+k)}{\Gamma(\Delta)\Gamma(k+1)}
        e^{-i(2k+\Delta)u},\qquad \operatorname{Im}u<0,
\end{align}
It is then useful to define  $a_k=\beta_+(2k+\Delta)$ and $c_k=\Gamma(\Delta+k)/[\Gamma(\Delta)\Gamma(k+1)]$. Poisson resummation then gives: 
\begin{align}
    G^+_{\mathrm{img},\epsilon}
    &=\frac{\beta_+^\Delta(2i)^\Delta}{2\pi}
        \sum_{k=0}^\infty c_k\sum_{\ell\in\mathbb Z}e^{i\ell\phi}I_{k\ell},\notag\\
    I_{k\ell}
    &=\int_{-\infty}^{\infty}\dd x\,e^{-i\ell x}
        (x-t+i\epsilon)^{-\Delta}e^{-ia_k(x+t-i\epsilon)}\notag\\
    &=e^{-i(\ell+2a_k)t-a_k\epsilon}
        \int_{-\infty}^{\infty}\dd y\,e^{-i(\ell+a_k)y}
        (y+i\epsilon)^{-\Delta}
\end{align}
Introducing the Schwinger parameter $s$, we can write 
\begin{align}
    (y+i\epsilon)^{-\Delta}
    &=\frac{i^{-\Delta}}{\Gamma(\Delta)}
        \int_0^\infty\dd s\,s^{\Delta-1}e^{isy-s\epsilon},\notag\\
    \int_{-\infty}^{\infty}\dd y\,e^{-ipy}(y+i\epsilon)^{-\Delta}
    &=\frac{2\pi i^{-\Delta}}{\Gamma(\Delta)}
        p^{\Delta-1}e^{-p\epsilon}\Theta_{\rm H}(p)
\end{align}
where $\Theta_{\rm H}$ is the Heaviside function. Note that the Fourier weight vanishes for $p=0$, and is nonzero for $p=a_k+\ell=p_{k\ell}>0$. This precisely reproduces the desired branch $Q>0$. Thus, we have
\begin{align}
    I_{k\ell}
    =\frac{2\pi i^{-\Delta}}{\Gamma(\Delta)}
        p_{k\ell}^{\Delta-1}\Theta_{\rm H}(p_{k\ell})
        e^{-i\omega_{k\ell}t-\omega_{k\ell}\epsilon}
\end{align}
Plugging it back into the propagator, the factor $2\pi$ as well as the phases cancel, and we obtain the expected mode sum:
\begin{align}
    G^+_{\mathrm{img},\epsilon}(t,\phi)
    =\sum_{k=0}^\infty\sum_{\substack{\ell\in\mathbb Z\\p_{k\ell}>0}}
        \frac{(2\beta_+)^\Delta p_{k\ell}^{\Delta-1}\Gamma(\Delta+k)}
            {\Gamma(\Delta)^2\Gamma(k+1)}
        e^{-i\omega_{k\ell}(t-i\epsilon)+i\ell\phi}
\end{align}
Taking $\epsilon\to0^+$ then reproduces (\ref{extremalpropagator}) including its normalization factors. The other branch can be simply obtained by $G^-(t,\phi)=G^+(-t,-\phi)$.

\bibliographystyle{jhep}
\bibliography{refs}

\end{document}